\documentclass{emulateapj}
\usepackage{amsmath, amsthm, amssymb, ifpdf}

\ifpdf
\else
\fi

\newcommand{\etal}{{\it et al.}}
\newcommand{\eg}{{\it e.g.,}}
\newcommand{\ie}{{\it i.e.,}}
\newcommand{\chisq}{{$\chi^{2}$}}
\newcommand{\msig}{{M$_{\rm BH} - \sigma_{*}$}}
\newcommand{\mlum}{{M$_{\rm BH} - $L$_{\rm bulge}$}}
\newcommand{\mbh}{{M$_{\rm BH}$}}
\newcommand{\ssig}{{$\sigma_{*}$}}

\begin{document}

\title{Black Hole - Bulge Relationship of Post-Starburst Quasars at $z\sim0.3$}

\author{Kyle D. Hiner\altaffilmark{1}, Gabriela Canalizo\altaffilmark{1}, Margrethe Wold\altaffilmark{2}, Michael S. Brotherton\altaffilmark{3}, Sabrina L. Cales\altaffilmark{3}}

\altaffiltext{1}{Department of Physics and Astronomy, University of California, Riverside, CA 92521, USA; email: kyle.hiner@email.ucr.edu, gabriela.canalizo@ucr.edu}
\altaffiltext{2}{Niels Bohr Institute Dark Cosmology Centre, University of Copenhagen, Copenhagen, Denmark; email: mwold@dark-cosmology.dk}
\altaffiltext{3}{Department of Physics and Astronomy, University of Wyoming, Laramie, WY; email: mbrother@uwyo.edu, scales@uwyo.edu}

\begin{abstract}

The \msig~relation has been studied extensively for local galaxies, but to date there have been scarce few direct measurements of stellar velocity dispersions for systems beyond the local universe. We investigate black hole and host galaxy properties of six ``post-starburst quasars'' at $z\sim0.3$. Spectra of these objects simultaneously display features from the active nucleus including broad emission lines and a host galaxy Balmer absorption series indicative of the post-starburst stellar population. These are the first measurements of \ssig~in such objects, and we significantly increase the number of directly-measured non-local objects on the \msig~diagram. The ``post-starburst quasars'' of our sample fall on or above the locally defined \msig~relation, a result that is consistent with previous \msig~studies of samples at $z>0.1$. However, they are generally consistent with the \mlum~relation. Futhermore, their location on the Faber-Jackson relation suggests that some of the bulges may be dynamically peculiar. \\

\end{abstract}
 

\singlespace

\section{Introduction}

The \msig~relation \citep{FM00,Gebhardt00,Gultekin09,Woo10} has been extensively studied in both active and inactive galaxies at low redshifts. The existence of this and other black hole - host galaxy scaling relations \citep{Magorrian98, McLureDunlop02} implies that the growth of black holes is intimately connected with that of the host galaxies. However, the mechanism by which this growth is regulated is still an open matter. 

One proposed mechanism is through black hole accretion feedback. \citet{DiMatteo05} show that only a modest amount of the accretion energy is required to shut down star formation in the host galaxy. Their simulations of galaxy mergers reproduce the local \msig~relation with only $\sim5$\% of the AGN luminosity being thermodynamically coupled to the surrounding gas. On the other hand, the need for accretion feedback is not clear. \citet{Peng07} argues that mergers between galaxies (major and minor) naturally leads to a tight log-linear M$_{\rm BH}$-M$_{\rm bulge}$ relation. To investigate the co-evolution of black holes with their hosts, it is necessary to measure the relation at higher redshifts.

\citet{Robertson06} study the theoretical evolution in the \msig~relation by simulating galactic mergers and including both supernova and black hole feedback. They predict that the slope of the relation remains roughly constant out to redshift $z=6$, while the scaling of the relation slightly decreases due to an evolving Faber-Jackson relation. The effect of this would be that $\sigma_{*}$ increases for a given M$_{\rm BH}$ as a function of redshift. Conversely, \citet{Croton06} predicts that the M$_{\rm BH}$-M$_{\rm bulge}$ relation evolves in the opposite sense, \ie~black hole mass increases with redshift for a given bulge mass.

Observational investigations into the evolution of the \msig~relation face difficulties. The sphere of influence of the black hole cannot be resolved at high redshifts, thus the methods using stellar or gas dynamics \citep{KR95} to measure M$_{\rm BH}$ in the local universe are unavailable. Alternatively, one can use the virial method to estimate black hole masses, which invokes AGN broad line widths and the R$_{\rm BLR}$-L$_{5100}$ relation \citep{Kaspi00,Bentz09a}. However, in type-1 active galactic nuclei (AGN), the active nucleus often outshines the host, drowning out stellar features in the spectra. Despite this effect, \citet{Woo06, Woo08} studied a sample of Seyfert galaxies and found that black holes at $z=0.36$ and $z=0.57$ were over-massive for their given stellar velocity dispersions compared to the local relation. \citet{Canalizo12} avoid the problem of an over-powering AGN continuum by studying dust-reddened quasars between $0.14 < z < 0.37$, which have spectra that show both active nucleus and host galaxy features. They find a result similar to Woo \etal~in that reddened quasars fall above the local \msig~relation. The red quasars of Canalizo \etal~also fall above the \mlum~relation, supporting the notion that the black holes are over-massive compared to their bulges.

Other recent studies have discussed regimes in which the local inactive galaxy relation may not hold. \citet{Graham11} calibrate the \msig~relation using elliptical, barred and unbarred galaxies and find that the slope of the \msig~relation can vary according to morphology. Conversely, \citet{Beifiori11} and \citet{Vika11} find no correlation between black hole mass and S\'{e}rsic index, {\it n}.  At the low-mass end of the \msig~relation, \citet{Xiao11} find little difference in the relation as defined by barred and unbarred Seyfert 1 galaxies. Xiao \etal~also state that inclination angle of a galactic disk components may increase the scatter of the \msig~relation. This is consistent with the results of \citet{Bennert11}, who find that \ssig~can be biased to larger or smaller values by the disk component of the galaxy. Recently, \citet{McConnell11b, McConnell11a} showed that brightest cluster galaxies have black holes that are significantly more massive than predicted from the \msig~relation given their velocity dispersions. 

In this paper we exhibit the first results of our investigation of the \msig~relation using a sample of AGN whose host galaxies contain luminous post-starburst stellar populations between redshifts $0.2 < z < 0.4$. \citet[][(in preparation)]{Brotherton12} catalog a subset of quasars with strong Balmer absorption series visible in their spectra. These ``post-starburst quasars'' (PSQs) are a convenient case for the study of the \msig~relation at higher$-z$, because spectra of these objects simultaneously show the broad emission lines distinctive of an active black hole as well as strong stellar absorption lines indicative of a post-starburst stellar population in the host galaxies. The moniker ``post-starburst quasars'' is not a definitive classification of the objects' luminosities. While some are luminous enough to be considered quasars, others are not. To be consistent with previous work \citep{Cales11, Brotherton10}, we will continue to refer to the sample as ``post-starburst quasars''.

In this paper we present the stellar velocity dispersions ($\sigma_{*}$) and black hole masses (M$_{\rm BH}$) of six post-starburst quasars as measured from spectroscopy performed with the Low Resolution Imaging Spectrograph (LRIS) on Keck I. Each of these PSQs were observed with $HST$ ACS F606W as part of a snapshot program (10588, PI Brotherton, M.). These images and a detailed morphological study of the host galaxies are presented in \citet{Cales11}. In our analysis we rely on photometry reported by Cales \etal, and we adopt the same cosmology that they use: a flat universe with H$_{\rm o} = 73$ km s$^{-1}$, $\Omega_{\rm M} = 0.27$, and $\Omega_{\Lambda} = 0.73$. The paper is organized as follows: we discuss the sample and observations in section 2, describe our spectral fitting code and $\sigma_{*}$ measurements in section 3, and discuss virial black hole mass calculations in section 4. We present our main results and discussion of possible biases in sections 5 and 6, respectively.

\section{Sample and Observations}

We present results for six post-starburst quasars from an overall sample of 609 PSQs selected from the Sloan Digital Sky Survey (SDSS) DR3 \citep{Abazajian05} that show both strong Balmer series absorptions and broad emission lines. The full details of the sample selection are described in \citet[][(in preparation)]{Brotherton12}, but see also \citet{Cales11} and \citet{Brotherton10}. Briefly: objects in the sample have total Balmer equivalent widths H$_{total} =$ EW$_{\rm H\delta}$ $+$ EW$_{\rm H8}$ $+$ EW$_{\rm H9}$ $ > 2$ \AA. H$\epsilon$ was excluded from this measurement, because it can be blended with Ca II. The Balmer break flux ratio was defined by two 100 \AA~regions starting at restframe wavelengths 3740 and 3985 \AA, and the selection criterion was set at $f_{3985 \rm \AA}/f_{3740 \rm \AA}  > 0.9$. The continuum S/N in the SDSS spectra were $>8$ as measured between 4150 and 4250 \AA.

We observed six post-starburst quasars of the sample using the Low Resolution Imaging Spectrometer \citep{Oke95,Rockosi10} at Keck I in September 2010. Conditions were excellent with $\sim0\farcs6$ seeing. We employed both the red and blue side cameras of LRIS to obtain broad spectral coverage. With this coverage we are able to observe the Balmer series (except for H$\alpha$ and H$\beta$) of our science targets. We observed several individual stars to construct host galaxy templates on nights in May 2010 and September 2010 using the same instrumental setup as for our science targets.

\begin{figure}
\epsscale{1.0}
\plotone{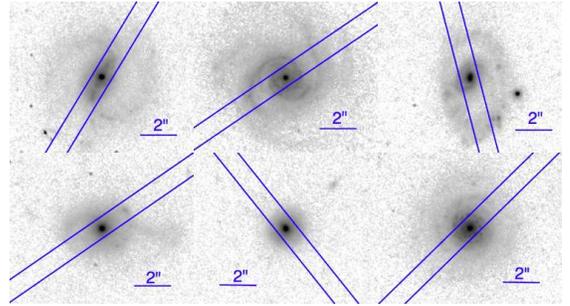}
\caption{HST ACS F606W images of our sample (Proposal ID 10588; PI: Brotherton, M.). North is up and East is to the left in each of the images. The LRIS slit was placed at parallactic angle with a 1$\arcsec$ width. {\it Top}: $0030-1035$, $0057+0100$, and $0237-0101$. {\it Bottom}: $2102+0005$, $2118+0056$, and $2306-0100$.\\}
\label{img}
\end{figure}

We observed all of the sample targets with the same setup with LRIS blue and red sides. On the blue side we used the 600/4000 grism, and on the red side we used the 1200/7500 grating. This setup covers observed frame 3100 \AA~to 6250 \AA. Because of the small dispersion in redshift of our science targets, this resulted in slightly different rest-frame wavelength coverage. With this setup we observed the quasar continuum and Mg II emission line on the blue side and the Balmer series on the red side. For one object, SDSS $0237-0101$, we also observed a second red region out to observed frame 7800 \AA. For each target we placed a 1\arcsec\ observing slit across the nucleus of the object and oriented it according to the parallactic angle. In Fig.~\ref{img}, we show $HST$ ACS F606W images of our sample with the LRIS slit overplotted (proposal 10588; PI: Brotherton, M.). Table \ref{observations} shows the redshift, exposure times, SDSS $r$ mag, the aperture size used to extract the spectra, and the morphology of each science target.

\begin{table*}
\caption{Observations \label{observations}}
\begin{tabular}{clccccc}
\hline
N & Object & redshift & exp time (s) & r [mag] & ap radius (\arcsec) & morphology\\
\hline
1 & SDSS $003043.59-103517.6$ & 0.294 & 1200 & 18.41 & 2.52 & bulge+disk\tablenotemark{1}, barred-spiral \\
2 & SDSS $005739.19+010044.9$ & 0.253 & 1200 & 17.91 & 2.28 & bulge+disk\tablenotemark{1}, spiral \\
3 & SDSS $023700.31-010130.4$ & 0.344 & 2400 & 18.19 & 0.22 & bulge+disk, spiral \\
4 & SDSS $210200.42+000501.9$ & 0.329 & 2400 & 18.66 & 0.98 & elliptical \\
5 & SDSS $211838.12+005640.6$ & 0.384 & 1200 & 18.36 & 0.44 & elliptical \\
6 & SDSS $230614.18-010024.4$ & 0.267 & 1200 & 18.76 & 0.26 & bulge+disk\tablenotemark{1}\\
\hline
\end{tabular}
\tablenotetext{1}{Objects are in the face-on orientation.}\\
\tablecomments{September 2010 Observations. SDSS 0237-0101 was observed with two setups on the red side of LRIS to obtain broad spectral coverage. The blue side was exposed twice for 1200 seconds each for a total of 2400 seconds on the blue side, and 1200 seconds in each of the red side setups. The sixth column shows the extraction aperture radii in arc seconds. These are the bulge effective radii from \citet{Cales11}. The last column describes the basic morphology of the object.}
\end{table*}

We followed standard data reduction procedures using the NOAO package in IRAF. First we removed bias levels using the overscan region and cosmic rays from the 2D spectra using the L.A.Cosmic routine \citep{vDokkum01}, and corrected for flat fields. We derived a wavelength solution from Hg, Ne, Ar, Cd, and Zn arc lamps and shifted individual spectra taken throughout the night based on sky lines. We fit sky background with a low-order polynomial and subtracted the fit from the 2D spectra. Based on the measurements of \citet{Cales11}, we extracted each spectrum using an aperture width equal to twice the effective radius of the host's bulge componet (\ie~center $\pm$R$_{\rm eff}$). This is consistent with what has been done in previous studies \citep{Gebhardt00, Bennert11} and avoids added uncertainty from aperture corrections when measuring stellar velocity dispersions. We used three standard stars observed throughout the night at parallactic angle to derive a flux calibration with data from \citet{Massey88} and \citet{Massey90}. We also corrected for Galactic reddening using the extinction values of \citet{Schlegel98} calculated at sky positions by the NASA Extragalactic Database extinction calculator. We used the IRAF task {\it deredden}, which uses the extinction function of \citet{CCM89} with the input extinction values to correct the spectra. 

We recovered the spectral region from 3100 to 7250 \AA~with the 600 line mm$^{-1}$ grism and 1200 line mm$^{-1}$ grating blazed at 7500 \AA~for each template star. The S/N of the template stars spectra was $\ge$1000. Table \ref{stars} lists the stars and their spectral types used to create host galaxy templates in this project. We conducted the same data reduction procedures on the observed template stars as on the post-starburst quasars with the additional step of correcting for observed radial velocities.

\begin{table}
\caption{Template Stars \label{stars}}
\begin{tabular}{lc}
\hline
Star & Sp Type \\
\hline
HD 93700 & A0 \\
HD 94980 & A3 \\
BD +182402 & A5 \\
HD 95002 & F0 \\
BD +182401 & F5 \\
HD 95958 & F8 \\
HD 97127 & G0 \\
HD 95676 & G5 \\
HD 184850 & G8 \\
HD 95437 & K0 \\ 
HD 230519 & K2 \\
HD 347822 & K2 \\
HD 347838 & K5 \\ 
HD 96761 & K5 \\
HD 347857 & K7 \\
BD +212255 & M0 \\
GSC $01562-00224$ & M8 \\
\hline\\
\end{tabular}
\tablecomments{In the cases where the same spectral type is repeated we used an average of the two stars for the host galaxy templates.}
\end{table}

The blue and red sides of LRIS have different spectral resolutions. The Balmer series fell on the blue side of LRIS for the template stars, but was shifted to longer wavelengths for the quasars in the sample. Before combining our spectra and fitting, we convolved the red side of the quasar spectra with a Gaussian to match the resolution of the blue side. We achieved an intrinsic resolution of $\sigma$ = 90 km s$^{-1}$ for the lowest redshift object (SDSS $0030-1035$; $z$=0.253) at rest-frame 3839 \AA. We achieved resolution of $\sigma$= 82 km s$^{-1}$ for the highest redshift object (SDSS $2118+0056$; $z$=0.384) at rest-frame 3475 \AA.

\section{Stellar Velocity Dispersion}

\subsection{Fitting Code}

To measure the stellar velocity dispersion of each host galaxy, we follow a direct fitting method that has been used extensively in previous studies. \citet{Barth03} use the direct fitting method to fit the spectra of BL Lac objects. More recently it was employed by \citet{Xiao11} to fit a sample of Seyfert 1 host galaxies, and by \citet{Greene10b} to fit the spectra of H$_2$O megamasers. \citet{Canalizo12} use the method to fit the host galaxies of dust-reddened quasars. 

We model the spectra with contributions from both the host galaxy and the active nucleus. The host template is constructed with a composite of stellar spectra taken with the same instrumental setup (see \S~3.2 for details) and broadened with a Gaussian in order to simulate the stellar velocity dispersion of the host galaxy. We include a power-law continuum attributed to the intrinsic active nucleus spectrum. We also include a Legendre polynomial that characterizes non-stellar emission in the spectra, including Fe emission and reddening effects. The polynomial makes no assumption regarding the reddening law (\eg~SMC). Thus, the full model function used in the fitting takes the form:

\begin{equation}
M(x)=\left\{ \left[ T(x) \otimes G(x) \right] + C(x) \right\} \times P(x)
\end{equation}

\noindent where T(x) is the host galaxy stellar template, G(x) is a Gaussian, C(x) is a power-law continuum (log($f$) = c$_1$ + c$_2$*log($\lambda$)), and P(x) is a third order Legendre polynomial. The fitting is performed in velocity space (log$\lambda$). There are a total of nine free parameters: the Gaussian velocity dispersion, an offset velocity between the model and the data, the power-law continuum scaling and slope, the four coefficients of the Legendre polynomial, and the relative flux contribution of an ``old'' and ``young'' host galaxy population. The ``young'' population template represents the post-starburst component separate from the older population of the host galaxy (see below).

The fitting code uses the amoeba routine from Numerical Recipies (written for IDL by W. Thompson) to minimize the \chisq~function. The inputs of the amoeba routine are the observed galaxy, the template model to fit, and the error for each datum in the observed galaxy spectrum. We created artificial errors to weight our data, focusing the fitting region on the Balmer series. To block emission lines from the fit, we set the pixel errors to a value several orders of magnitude larger than the error on the unblocked regions. Thus these emission lines do not contribute to \chisq, and are ignored during the minimization.

\subsection{Templates}

We must be careful in defining a host galaxy template that correctly matches the stellar population of the galaxies. A template mis-match from the stellar population of the host galaxy may cause an incorrect measure of the velocity dispersion. Mismatch between the host stellar population and the template composition can be a source of error in the direct fitting method \citep{GH06b, Barth02}. Furthermore, the Balmer lines of early type stars are intrinsically broadened by the combined effects of turbulence in the stellar atmosphere, axial rotation, and pressure broadening of the energy levels within the hydrogen atoms. This intrinsic broadening may bias our measured \ssig~if we were to use a host template that did not include early type stars. As long as the templates used in the direct fitting include early type stars with luminosity weighted proportions that mimic the galaxy stellar population, then the broadening function simulates the kinematics of the host galaxy, not the intrinsic line widths. 

We set out to create host galaxy templates that matched the stellar populations in the host galaxies, including the early-type stars. To do this we collected spectra of a set of stars ranging across 15 spectral types from A0 through M8. The spectral range covers $3100-7200$ \AA. 

We simulate single-epoch starburst models by determining the luminosity weighted proportions of each spectral type that contribute to individual models. We produced a series of templates to compare to the instantaneous burst models of Charlot \& Bruzual (private communication, see also \citealt{Bruzual07}). The series begins with weights favoring early type stars to simulate a young stellar population, then progresses by removing the earliest spectral types and weighting the rest of the stars accordingly. When compared with the starburst models of Charlot \& Bruzual, this simulates a range of starburst ages by ``evolving'' early type stars out of the template.

We compared the Charlot \& Bruzual models to each template produced in the series as well as to the individual stars that were included in the templates. Using the \chisq~statistic as an indicator of the goodness-of-fit, we identified nine unique matchings between the Charlot \& Bruzual models and our templates. We note that we only calculated \chisq~across the spectral fitting region 3600 - 4800 \AA~in order to obtain the best match in the region of interest, but the templates generally matched the Charlot \& Bruzual models well at longer wavelengths as well. 

Thus, we produced spectral templates that match stellar populations of ages 321 Myr, 404 Myr, 570 Myr, 640 Myr, 718 Myr, 904 Myr, 4.0 Gyr, 6.0 Gyr and 9.0 Gyr. The 321 Myr model was best matched with a single A0 star. The 404 Myr model was best matched with a single F0 star. The 6 Gyr and 9 Gyr models were best matched by G5 and G8 stars, respectively. The remaining models were best fit by templates that included all spectral types from our library with proportions weighted by luminosity toward early type stars. 

\subsection{Fitting Procedure}

While the host galaxies show a strong post-starburst component, they also contain an older stellar population. The observed line widths can be fitted using the Gaussian function, but the line strengths, ratios, and continuum shape will be determined by the age of the post-starburst population and its relative contribution to the overall flux. For this reason, we adopt a procedure that tests each of the templates that we created and by doing so determine an approximate age of the starburst population. The best fitting templates were combinations of the 9 Gyr template (a single G8 star) with templates that were $<1$ Gyr old. We allowed the relative contribution of the ``old'' and ``young'' populations to be a free parameter. The final values of the parameters can be sensitive to their initial values. This is because the \chisq~value can converge on a local minimum within the parameter space. To ensure we obtain a robust result that converges on the global minimum, we re-fit the spectra after perturbing the input parameters. We present the measured velocity dispersions (\ssig), luminosity-weighted fraction of the starburst component, 95\% confidence interval of \ssig, and the age of the younger starburst template in Table \ref{sigmas}.

\begin{table*}
\caption{Best Fit Parameters \label{sigmas}}
\begin{tabular}{lccccc}
\hline
& & & & 95\%  & younger population \\
N & Object & $\sigma_{*}$ & SB & low-high & age \\
\hline
1 & SDSS $003043.59-103517.6$ & $73$ & 99 & $44-79$ & 718 Myr \\
2 & SDSS $005739.19+010044.9$ & $134$ & 33 & $126-146$  & 321 Myr \\
3 & SDSS $023700.31-010130.4$ &  $214$ & 79 & $209-233$ & 640 Myr \\
4 & SDSS $210200.42+000501.9$ & $162$ & 86 & $154-166$ & 404 Myr \\
5 & SDSS $211838.12+005640.6$ & $134$ & 63 & $118-137$ & 404 Myr \\
6 & SDSS $230614.18-010024.4$ & $54$ & 86 & $34-61$ & 570 Myr \\
\hline\\
\end{tabular}
\tablecomments{The velocity dispersions are shown under the column labeled \ssig~and have units  km s$^{-1}$. Velocity dispersions reported here have not been corrected for the aperture. The column labeled SB are the fractional contribution of the star-burst component to the total flux shown as percentages. The lower and upper limits of the 95\% confidence interval is shown with units  km s$^{-1}$. In the last column we report the age of the template used to represent the young stellar population.}
\end{table*}

To determine the 95\% confidence interval of our $\sigma_{*}$ measurements, we re-fit the spectra holding $\sigma_{*}$ constant accross the range $\pm30$ km s$^{-1}$ from the best fit $\sigma_{*}$ in intervals of 5 km s$^{-1}$. We inspected the resulting \chisq~versus $\sigma_{*}$ plot to ensure that the measured $\sigma_{*}$ was indeed the value that minimized \chisq, and we fit a parabola through the values in order to define more precisely the confidence interval. The 95\% confidence interval is defined as the range in \ssig~that increases the \chisq~value by 4 \citep{Press07}. 

\begin{figure*}
\epsscale{1.0}
\plotone{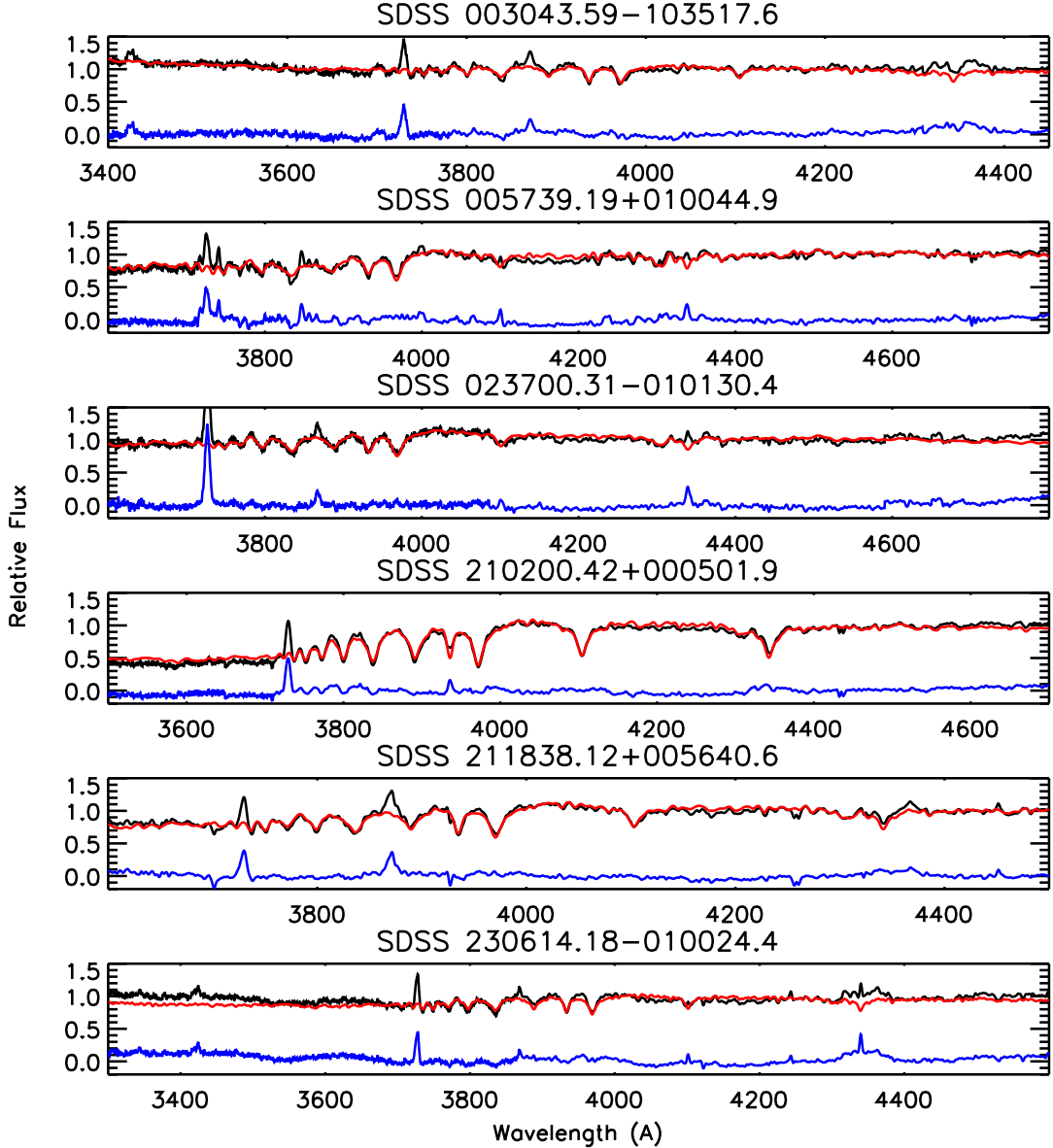}
\caption{Spectral fitting results. The data are plotted in black, the best fit model in red, and the redsidual is plotted in blue. Each spectrum is plotted in the rest frame.\\}
\label{fitresults}
\end{figure*}

Figure~\ref{fitresults} shows the best fits obtained for our targets. The observed spectra are plotted in black. The best fit model is overplotted in red, and the residuals are plotted in blue.

\begin{figure}
\epsscale{1.0}
\plotone{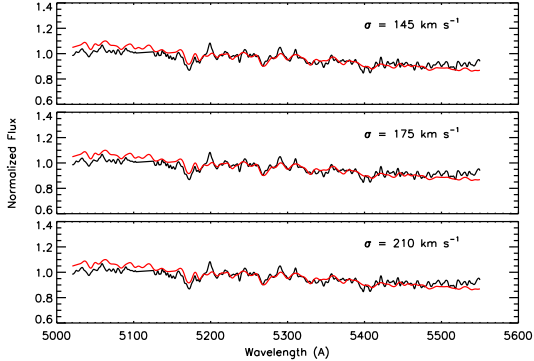}
\caption{Spectral fitting results for the \ion{Mg}{1b} region of SDSS $0237-0101$. We have plotted the best fits for three values of \ssig: 145, 175, and 210 km $^{-1}$. The data are plotted in black, the best fit model in red.\\}
\label{mgfit}
\end{figure}

We obtained an additional longer wavelength spectrum that extends to rest frame 5800 \AA~for the object SDSS $0237-0101$. We fit the spectral region around \ion{Mg}{1b} with the expectation that features from the underlying older stellar population would be dominant in this region. The exact region was from $5020-5550$ \AA~with one emission line at 5198 \AA~blocked from the fit. Fitting this region can be complicated due to [Mg/Fe] correlation with velocity dispersion \citep{Jorgensen99,Wold07}. However, we found identical results between fits conducted with and without the \ion{Mg}{1b} absorption blocked from the fit. Despite some effort in matching the templates to the host galaxy and using various weighting schemes, we were unable to derive a robust measure of \ssig~from this region. The best fit dispersion using this region was 155 (+19, $-$18) km s$^{-1}$, which is smaller than the dispersion measured from the Balmer region, i.e., 214 (+19, $-$5) km s$^{-1}$). However, we were able to fit \ssig~values that ranged from $\sim145\sim210$ km s$^{-1}$ while still maintaining visually similar fits. In Figure~\ref{mgfit}, we show example fits for this region using \ssig~values of 145 km s$^{-1}$, 175 km s$^{-1}$, and 210 km s$^{-1}$. As in Figure~\ref{fitresults}, the data are plotted in black, and the models are plotted in red. The residuals for the three fits shown are similar to one another. These fits indicate that there are likely model degeneracies that are affecting the determination of \ssig~in this region.

\section{Black Hole Mass}

Black hole masses can be estimated using the virial method under the assumption that the broad line region (BLR) gas is virialized. To calculate the virial parameter, we require measurements of the gas velocity and orbital radius. The velocity is, in principle, easily measured from broad line widths (\eg~Mg II, H$\beta$, H$\alpha$). Furthermore, reverberation mapping studies have shown a correlation between the radius of the BLR (R$_{\rm BLR}$) and the continuum luminosity of the active nucleus \citep{Kaspi00, Bentz09a}. Thus, the observables to employ the virial method are the line width and continuum luminosity.

To convert the virial parameter ($VP=R_{\rm BLR}\Delta v^2 / G$) to a black hole mass, one must include an additional {\it f}-factor, which accounts for unknown geometries in the broad line region. \citet{Woo10} have measured this factor by scaling the virial parameter of reverberation mapped AGN to match the \msig~relation, and found {\it f} $= 5.2$. This is consistent with values found by previous authors \citep{Onken04} that are about 1.8 times higher than an {\it f}-factor assuming an isotropic BLR velocity field. 

In the following sections we describe how we measured the broad line widths and continuum luminosities used in the virial black hole mass calculation. We measure the H$\alpha$ FWHM, and the Mg II $\lambda\lambda 2796,2803$ FWHM to use as BLR gas velocities. We measure the 5100 \AA~AGN continuum luminosity to use as a proxy for the BLR radius. In the following we will refer to $\lambda$L$_{\lambda}$ as L$_{3000}$ or L$_{5100}$ for $\lambda = 3000$ \AA~and $\lambda = 5100$ \AA, respectively. Table \ref{observables} displays the relevant measurements. Table~\ref{masses} shows the mass estimates and Eddington ratios.

\begin{table*}
\caption{Line Widths and Monochromatic Luminosities \label{observables}}
\begin{tabular}{lcccccccccc}
\hline
& & H$\alpha$ & \multicolumn{2}{c}{Mg II} & L$_{3000}$ & L$_{5100}$ \\
N & Object & FWHM & $\sigma_{intr}$ & FWHM$_{intr}$ & 3000 \AA & 5100 \AA\\
\hline
1 & SDSS $003043.59-103517.6$ & 4981 $\pm$ 428 & 1563 & 3066 & 4.7$^{+1.7}_{-1.2}$ & 1.5$^{+0.6}_{-0.4}$ \\ 
2 & SDSS $005739.19+010044.9$ & 4300 $\pm$ 210 & 1871 & 3464 & 0.36$^{+0.07}_{-0.06}$ & 0.30$^{+0.08}_{-0.06}$ \\ 
3 & SDSS $023700.31-010130.4$ & 7125 $\pm$ 770 & 2357 & 4855 & 0.39$^{+0.2}_{-0.1}$ & 0.41$^{+0.2}_{-0.1}$ \\ 
4 & SDSS $210200.42+000501.9$ & 6132 $\pm$ 315 & \nodata & \nodata & 0.58$^{+0.08}_{-0.07}$ & 0.79$^{+0.06}_{-0.06}$ \\ 
5 & SDSS $211838.12+005640.6$ & 6303 $\pm$ 501 & 2278 & 2913 & 1.4$^{+0.4}_{-0.3}$ & 1.7$^{+0.3}_{-0.5}$ \\ 
6 & SDSS $230614.18-010024.4$ & 3704 $\pm$ 719 & 1857 & 2468 & 1.1$^{+0.7}_{-0.4}$ & 1.2$^{+1.2}_{-0.4}$ \\ 
\hline\\
\end{tabular}
\tablecomments{Widths are presented with units km s$^{-1}$, and luminosities are presented with units {\sf x}10$^{44}$ erg s$^{-1}$. The H$\alpha$ FWHM have been corrected for the instrumental resolution of the SDSS using R$=2000$. The Mg II widths have been corrected for the instrumental resolution of LRIS. Instrumental resolution had a small effect on the widths compared to the errors. Errors on the luminosities were determined by varying the fitted slope by $\pm0.1$.} 
\end{table*}

\subsection{H$\alpha$}

The H$\beta$ broad emission line is often used to calculate black hole masses in single-epoch spectra of AGN. However, in our sample, the absorption-line spectrum of the host galaxy can significantly contaminate the H$\beta$ emission line. Often the broad H$\beta$ line is completely unobserved, so H$\beta$ emission is not suitable for measuring \mbh. On the other hand, H$\alpha$ is much less affected by the stellar absorption line than H$\beta$, and broad H$\alpha$ is always observed in the spectra of the targets. 

To calculate black hole masses we adopt the relation from \citet{Greene10a}: 

\begin{equation*}
{\rm M_{BH}} = (9.7 \pm 0.5){\sf x}10^6(\frac{\rm {L_{5100}}}{10^{44} {\rm~erg~s^{-1}}})^{0.519 \pm 0.07}
\end{equation*}
\begin{equation}
{\sf x}(\frac{{\rm FWHM_{H\alpha}}}{10^3~{\rm km~s^{-1}}})^{2.06 \pm 0.06}~{\rm M_{\odot}}
\end{equation}

\noindent Greene \etal~use a virial coefficient from \citet{Onken04} ({\it f}$=5.5$), which is 1.8 times higher than the assumption of isotropic random motions. Their equation was originally determined by \citet{GH05b}, who derive a conversion between H$\alpha$ and H$\beta$ FWHM. Thus, the above equation converts the H$\alpha$ FWHM to that of H$\beta$, and uses an {\it f}-factor appropriate for H$\beta$.

\citet{Woo10} measure the \msig~relation for the reverberation mapped sample of the Lick AGN Monitoring Project \citep{Walsh09, Bentz09b, Bentz10}. Woo \etal~determine the geometric factor for {\it f}$=5.2^{+1.2}_{-1.3}$ by scaling the determined H$\beta$ virial products to the local \msig~relation for inactive galaxies determined by \citet{Gultekin09}. Because the relation from \citet{Greene10a} uses the same virial coefficient as found by Woo \etal, any differences we find between our sample and that of Woo \etal~will not be due to differences between {\it f}-factors.

\begin{figure*}
\epsscale{1.0}
\plotone{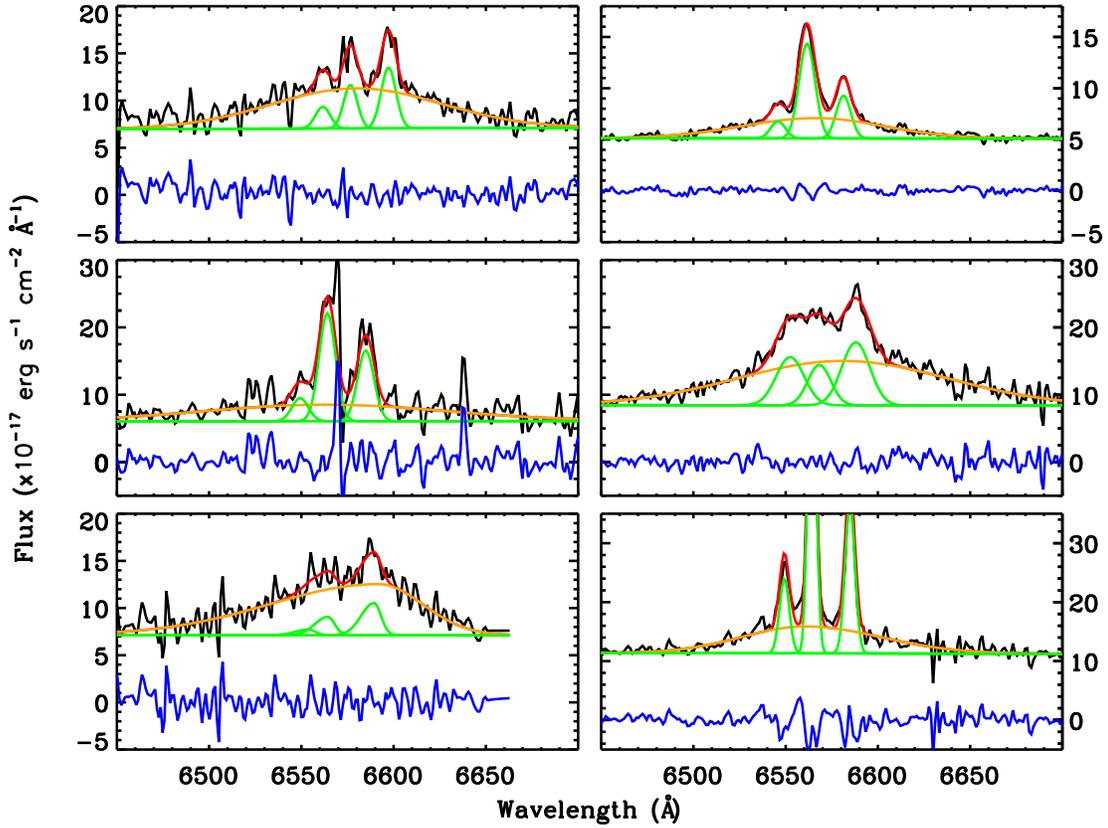}
\caption{Broad- and narrow-line decompositions of the H$\alpha$ region. The SDSS DR7 spectra are plotted in the restframe in black. The models of the narrow emission lines and continuum are plotted in green, and the H$\alpha$ broad emission line component is plotted in orange. The overall fit is plotted in red. The residuals are plotted in blue. The objects in the left column from top to bottom are $0030-1035$, $0237-0101$, and $2118+0056$. The objects on the right are $0057+0100$, $2102+0005$, and $2306-0100$.}
\label{Halpha}
\end{figure*}

To measure the widths of the H$\alpha$ broad emission line we used the task {\it specfit} in the STSDAS package of IRAF \citep{Kriss94}. Our LRIS spectroscopy does not cover the H$\alpha$ region, so we measure the width from the SDSS spectra \citep[DR7; ][]{Abazajian09}. The task {\it specfit} minimizes the \chisq~using a simplex method and returns the central wavelength, integrated flux, FWHM, and skew of each line. We fit the region from 6400 - 6700 \AA~using five components: a power law continuum, three Gaussians for the narrow line components of [N II] $\lambda$6548, H$\alpha$, and [N II] $\lambda$6583, and an additional Gaussian for the H$\alpha$ broad emission line. We held the relative positions of the [N II] lines constant for the fitting procedure, and constrained all the Gausians to be symmetric. We fit a narrow line model using [O III] 5007 on objects where the line was visible and scaled the model to the [N II] and H$\alpha$ narrow lines. The {\it specfit} task also returns errors on the fitted parameters, which are generally large due to the resolution and S/N of the SDSS spectra. We show the fits of the H$\alpha$ region for all objects in Fig.~\ref{Halpha}.

The stellar H$\alpha$ absorption line from the host galaxy is small compared to each of these components, and significantly more narrow than the broad emission line of the active nucleus. Using spectra and host galaxy fits for PSQs from \citet[][(in preparation)]{Cales12} that cover H$\alpha$, we determined that the host galaxy absorption component does not affect the measurement of the broad line FWHM. 

The data of \citet[][(in preparation)]{Cales12} cover the H$\alpha$ region for the object SDSS $0057+0100$, and have higher resolution and S/N ($\sim$28 pix$^{-1}$ on the continuum) than the SDSS spectrum (S/N$\sim$10 pix$^{-1}$ on the continuum). Following the same procedure, the FWHM measured using these data agrees within the errors of the FWHM measured from the SDSS spectrum. We adopt the width measured with the data from Cales \etal, because the S/N is an improvement over the SDSS spectrum. 

\subsection{Mg II}

As with the H$\alpha$ emission line, there are several prescriptions for calculating black hole masses from the Mg II $\lambda\lambda 2796,2803$ broad emission doublet \citep{McLure02, McGill08, Rafiee11, VO09}. Of these, the calibration from \citet{VO09} is based on the largest number of objects and is calibrated with L$_{5100}$ in addition to other continuum luminosities. As described in the next section, we consider the 5100 \AA~continuum luminosity to be better constrained than the 3000 \AA~luminosity, which is typically used in conjunction with the Mg II emission to calculate black hole masses. The formulation that we adopt from Vestergaard \& Osmer is:

\begin{equation}
{\rm M_{BH}} = 10^{6.96} (\frac{{\rm FWHM_{Mg II}}}{1000 {\rm km s^{-1}}})^2 (\frac{\rm {L_{5100}}}{10^{44} {\rm erg s^{-1}}})^{0.5}~{\rm M_{\odot}}
\end{equation}

\noindent According to Vestergaard \& Osmer, the uncertainty in the zero point of this relation is 0.55 dex. This is the dominant factor in the uncertainty on the black hole mass calculated from the Mg II emission. 

To measure the Mg II width, we first fitted and subtracted an Fe II emission template from the spectrum. Iron emission can contaminate the flux surrounding the Mg II emission and affect the measurement of the broad line width. We used the template from \citet{Vestergaard01}, who derived the template from the observed spectrum of the narrow-line Seyfert 1 galaxy I Zw 1. 

The template by \citet{Vestergaard01} only includes Fe emission that may contaminate the wings of the Mg II profile. To account for Fe emission that may contaminate the core of the line, \citet{Fine08} and \citet{Kurk07} add a constant flux under the Mg II core equal to 20\% of the average Fe flux at $2930-2970$ \AA. This is based on the theoretical Fe emission models of \citet{Sigut03}. To determine if the theoretical flux may have affected our Mg II width measurements, we subtracted the additional flux from the fit of SDSS $0237-0101$ and re-measured the width of the Mg II emission. There was no difference measured between the two methods, so we do not believe this small amount of theoretical flux has affected the measurements of Mg II widths. 

We used a fitting routine similar to the one that we used to determine stellar velocity dispersions to scale and broaden the Fe template before subtraction. The main differences between the way that we modeled the Balmer region and the Mg II region are that we replaced the host galaxy stellar template with the Fe template from \citet{Vestergaard01} and the fitted continuum includes contributions from both the active nucleus and host galaxy at these wavelengths. We adopted a power law to fit the continuum, as we expect that the host galaxy contribution to the flux at these wavelengths is comparatively small. We fit a total of five parameters: normalization and slope for a power law continuum, and the scaling constant, velocity dispersion ($\sigma_{Fe}$), and velocity offset for the Fe template.

\begin{figure*}
\epsscale{1.0}
\plotone{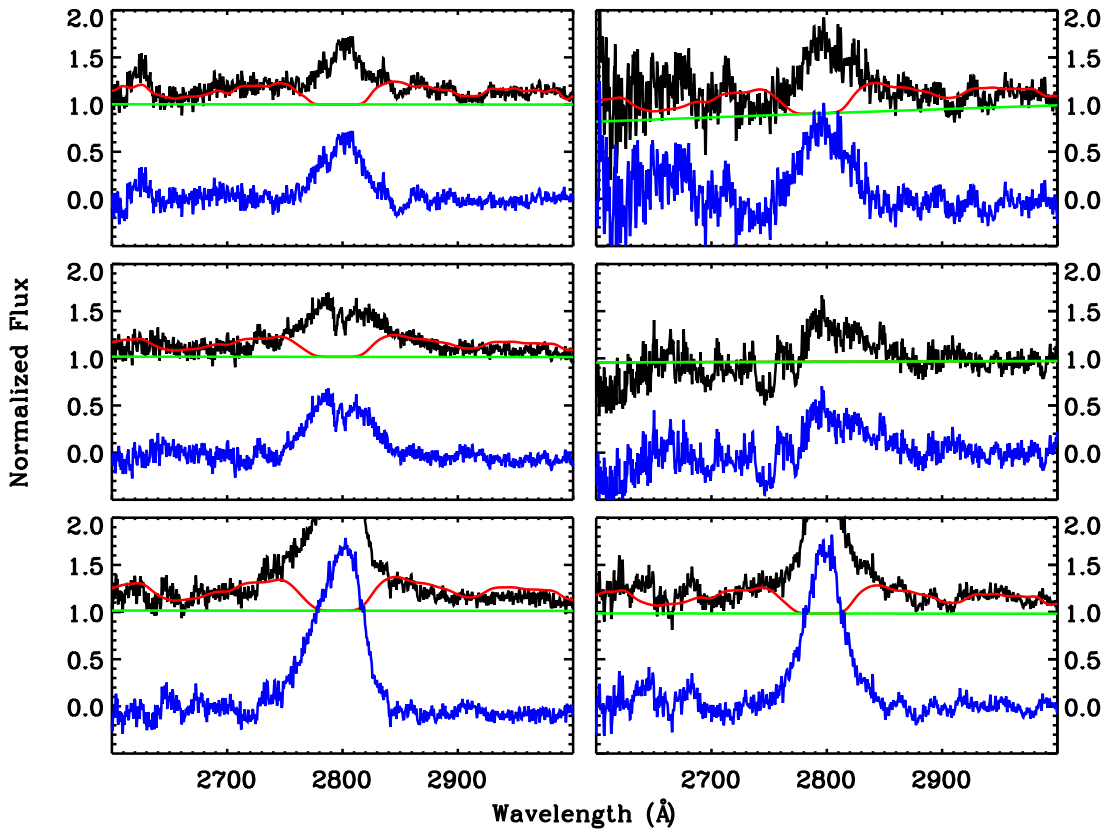}
\caption{Fe template fitting in the Mg II region. The LRIS data are plotted in black. The overall fitted Fe template is plotted in red, and the continuum is plotted in green. The residual spectrum (data - model) is plotted in blue. The objects are in the same order as Fig.~\ref{Halpha}}
\label{Feres}
\end{figure*}

We first estimated the continuum from a region relatively uncontaminated by Fe emission: $3010 - 3026$ \AA. The result was used as the continuum initial estimate for the main fitting routine. The Fe template does not include the Mg II emission line, so we blocked the feature from contributing to the \chisq~during the minimization. This region varied in size for each object depending on the strength of the emission. We allowed $\sigma_{Fe}$ to vary up to $2000$ km s$^{-1}$, but each fit produced a value smaller than 1000 km s$^{-1}$. The best fit model was then subtracted from the spectrum leaving a residual spectrum. We show Fe template fits to all objects in Fig.~\ref{Feres}.

Objects SDSS $0057+0100$ and SDSS $2102+0005$ were peculiar, and were not fit well by the Fe template (first and second object on the right side of Fig.~\ref{Feres}). The spectrum of SDSS $0057+0100$ shows some signs of absorption to the blue side of Mg II. However, the noise significantly increases at the edge of the spectrum, making the continuum level difficult to determine. It is possible that SDSS 0057+0100 is a low-ionization broad absorption line (LoBAL) quasar, but without a clear determination of the continuum level at shorter wavelengths, we hesitate to classify it as such. In order to fit the Fe template to this object, we blocked the wavelength region shortward of and including the Mg II emission ($<$ 2840 \AA) from the fit, allowing the overall fit to be determined by the longer wavelengths. Then we subtracted the entire fitted spectrum from the region before measuring the Mg II width. The spectrum of SDSS $2102+0005$ has better S/N on the blue side of Mg II than that of SDSS $0057+0100$, and the profile of Mg II makes it clear that the object is a LoBAL. Because the spectrum shows strong absorption over the blue wing of the Mg II line, we were unable to reliably measure a Mg II line width. We do not report a Mg II width measurement for SDSS $2102+0005$, but do include one for SDSS $0057+0100$.

\subsection{Continuum Luminosities}

The continuum luminosity is used as a proxy for the BLR radius in the virial method to measure black hole mass. We measure the continuum luminosity from the power-law continuum of our model. The power-law is multiplied by a Legendre polynomial to simulate reddening effects and non-stellar emission, and the product of the continuum and Legendre polynomial represents the reddened nucleus free of host contamination. In order to obtain the intrinsic continuum luminosities we follow a two step process: First, we calculate a scaling factor between the product of the continuum and Legendre polynomial fitted by our spectral code and the $HST$ F606W photometry of the nucleus, which has been deconvolved from the host galaxy \citep{Cales11}. The scale factor effectively flux calibrates our fitted spectrum to the $HST$ photometry. Then we apply the scaling factor to the fitted underlying power law continuum (without the Legendre polynomial).

We measure the monochromatic luminosities from the scaled power law. It is important to note that both 3000 \AA~and 5100 \AA~fall outside the spectral region that we modeled. Thus, the luminosites are extrapolations from our fitted quasar power-law continuum. We define uncertainty on the continuum luminosities by varying the slope of our intrinsic continuum by $\pm$0.1, which is the scatter in the average quasar spectral slope \citep{VandenBerk01}. The emission at 3000 \AA~has contributions from the active nucleus, the host galaxy, and iron. While we have a template that models the iron emission \citep{Vestergaard01}, attempting to separate the three components would include many free parameters and model degeneracies may become a significant issue. Furthermore, the photometry that we use to scale the fitted active nucleus spectrum is in the observed $HST$ F606W filter. Because 5100 \AA~is much closer to both our spectral fitting region and the F606W photometry, we consider L$_{5100}$ to be much better constrained than L$_{3000}$. We present both measured luminosities and their formal uncertainties in Table \ref{observables}. The true uncertainty for the L$_{3000}$ measurement is larger than the formal uncertainty that we have reported for the reasons noted above.

\section{Results}

In the previous sections we have presented our \ssig~and \mbh~measurements of six PSQs. We now examine the relationship between the black holes and their host bulges through the \msig~and \mlum~relations. We further examine the properties of the host galaxy bulges with the Faber-Jackson relation to determine if the bulges are dynamically peculiar compared to evolved systems in the local universe. By examining these three relations, we can determine if any observed offset is primarily due to over/under-massive BHs or dynamically perturbed bulges.

Using the H$\alpha$ and Mg II emission line widths, we calculated masses of the black holes in six post-starburst quasars and present the values in Table \ref{masses}. In five of the six objects, we measure larger masses based on the H$\alpha$ broad line than the Mg II broad line. We were unable to measure the \mbh~based on the Mg II broad line for SDSS $2102+0005$, because the line profile for this object is affected by an absorption trough on the blue wing, as described above. 

The relations from \citet{Greene10a} and \citet{VO09} are each calibrated to H$\beta$. In principle they should give consistent results, because the two prescriptions are measuring the same black hole mass, and both H$\alpha$ and Mg II should be reliable indicators of the BLR gas dynamics and \mbh. That is, according to \citet{GH05b}, the correlation between the H$\alpha$ FWHM and the H$\beta$ FWHM has a scatter on the order of $\sim0.1$ dex. Furthermore, \citet{VO09} report their calibration produces Mg II-based masses that are consistent with the H$\beta$ and C IV-based masses within $\sim0.1$ dex. However, the 1$\sigma$ scatter on the zero point of the Vestergaard \& Osmer relation is 0.55 dex. If we consider this large error associated with the Mg II-based calculation, then our calculated H$\alpha$ and Mg II masses are consistent. The discrepancy could arise if, at least for PSQs, the geometric {\it f}-factor were different between the emission line gas species. This is not unlikely, as magnesium has a lower ionization potential than hydrogen, and may arise from a different location within the BLR.

\begin{table*}
\caption{Black Hole Masses \label{masses}}
\begin{tabular}{lccccc}
\hline\\
& & M$_{\rm BH}$ & M$_{\rm BH}$ & L$_{bol}$/L$_{Edd}$ & L$_{bol}$/L$_{Edd}$ \\
N & Object & H$\alpha$ & Mg II & 5100 \AA & [O III] \\
\hline\\
1 & SDSS $003043.59-103517.6$ & 3.25$^{+3.0}_{-1.5}$ & 1.1 & 0.037 & 0.019 \\
2 & SDSS $005739.19+010044.9$ & 1.06$^{+1.1}_{-0.5}$ & 0.60 & 0.023 & 0.076 \\
3 & SDSS $023700.31-010130.4$ & 3.49$^{+4.0}_{-1.8}$ & 1.4 & 0.009 & 0.057 \\
4 & SDSS $210200.42+000501.9$ & 3.59$^{+3.2}_{-1.7}$ & \nodata & 0.018 & 0.013 \\
5 & SDSS $211838.12+005640.6$ & 5.70$^{+5.0}_{-2.6}$ & 1.0 & 0.024 & 0.027 \\
6 & SDSS $230614.18-010024.4$ & 1.56$^{+1.9}_{-0.8}$ & 0.60 & 0.060 & 0.062 \\
\hline\\
\end{tabular}
\tablecomments{Units of the black hole masses are {\sf x}10$^8$ M$_{\odot}$ and were calculated from the equations of \citet{Greene10a} and \citet{VO09}. Errors on the Mg II-based black hole masses are dominated by the calibration zero point error ($\sim0.55$ dex). Eddington ratios assume bolometric correction factors of 10 for the 5100 \AA~luminosities \citep{WooUrry02}, and 3500 for the [O III]$\lambda$5007 luminosities, respectively.}
\end{table*}

\subsection{The \msig~Relation}

\begin{figure*}
\epsscale{1.0}
\plotone{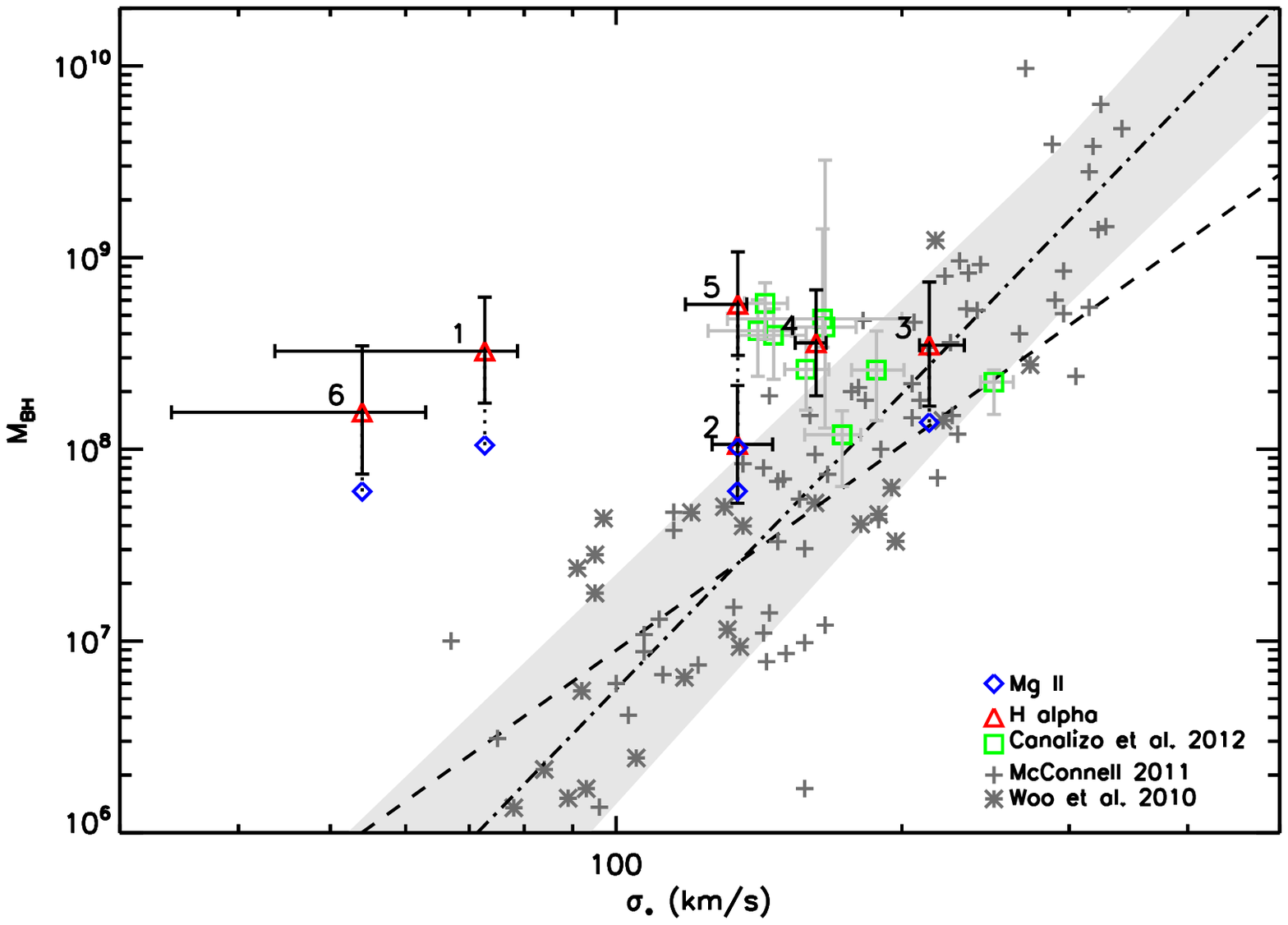}
\caption{Plot of our objects compared to the locally-defined \msig~relation. The relation by \citet{McConnell11b} is plotted as a dot-dashed line and grey plus symbols. The scatter is shown as a shaded grey region.The relation of reverberation mapped active galaxies by \citet{Woo10} is plotted as a dashed line with grey asterisks. We plot the red quasars from \citet{Canalizo12} in green squares. The black hole masses of the PSQs based on H$\alpha$ and Mg II are plotted as red triangles and blue diamonds, respectively.\\}
\label{msfig}
\end{figure*}

We have plotted the objects studied here on the \msig~diagram (Fig.~\ref{msfig}). For comparison we present the relation measured by \citet{McConnell11b}, who update the plot with compiled measurements from the literature. We also include reverberation mapped AGN from \citet{Woo10}. The best fit from McConnell \etal~is plotted as a dot-dashed line, and the fit from Woo \etal~is plotted as the dashed line. We plot the H$\alpha$ based black hole masses in red triangles. The scatter reported by both McConnell \etal~and Woo \etal~is $\sim0.43$ dex, and the H$\alpha$ based \mbh~all fall within or above this scatter with respect to either relation. Three of the black hole masses that were based on the Mg II measurement are consistent with this scatter.

The objects SDSS $0030-1035$ and SDSS $2306-0100$ appear significantly offset from the \msig~relation. Both of these galaxies were classified by \citet{Cales12} as face-on disks. The face-on disk presents a dynamically cold stellar component with small line-of-sight velocity dispersion, and \citet{Bennert11} showed that this orientation can bias the measured velocity dispersion toward smaller values. The sample of \citet{Woo06} contains four face-on disks, and the objects do have lower \ssig~values than the remainder of their sample. However, the magnitude of the bias for their objects is only $\sim30-50$ km s$^{-1}$ ($\Delta$\ssig~$\sim0.1$ dex), whereas the two objects here have offsets on the order of $\sim140$ km s$^{-1}$ ($\Delta$\ssig~$\sim0.6$ dex). Furthermore, as we show below, one of these objects is consistent with the Faber-Jackson relation. 

One other object in our sample was classified by \citet{Cales12} as a face-on disk (SDSS $0057+0100$), but it falls within the scatter of the \msig~relation. SDSS $0237-0101$ also has a bulge+disk morphology, but is not face-on and falls within the scatter of the \msig~relation. The last two objects (SDSS $2102+0005$ and SDSS $2118+0056$) are dominated by bulges and fall outside the defined 1$\sigma$ scatter of the \msig~relation.

\begin{figure}
\epsscale{1.0}
\plotone{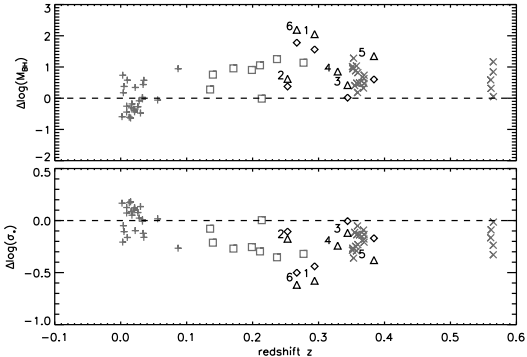}
\caption{The offset from the local \msig~relation \citep{Woo10} is plotted against the redshift of the objects. The PSQs from our sample are plotted twice with triangles representing the H$\alpha$-based \mbh~measurement and the diamonds the Mg II-based measurement (same as Fig.~\ref{msfig}). The red quasars of \citet{Canalizo12} are plotted as squares, and the Seyfert 1 galaxies from \citet{Woo06} and \citet{Woo08} are plotted as X symbols. The objects that define the local relation from \citet{Woo10} are plotted as plus symbols.\\}
\label{zevo}
\end{figure}

In Fig.~\ref{zevo} we show the offset of M$_{\rm BH}$ ({\it top panel}) and the offset of $\sigma_{*}$ ({\it bottom panel}) from the relation defined by \citet{Woo10} as a function of $z$. Previous studies have found an offset from the local \msig~relation, including \citet{Woo06,Woo08} and \citet{Canalizo12}. The objects in our sample span a redshift range that overlaps with the samples of Canalizo \etal~and Woo \etal. Furthermore, the trends seen by them appear to continue in our objects. As discussed by Canalizo \etal, at first glance this implies very recent evolution in the \msig~relation.

The average \mbh~offset for the sample of all six quasars with respect to the relation by \citet{Woo10} is 1.25 dex when determined from the H$\alpha$ broad line. If we exclude the object that does not conform to the Faber-Jackson relation (SDSS $2306-0100$, see below), then the average offset in \mbh~becomes 1.06 dex. This result is significantly higher than the scatter in the local \msig~relation, which is $\sim0.43$ according to both \citet{Woo10} and \citet{McConnell11b}. The offset we measure is slightly larger than the average offsets of red quasars found by \citet{Canalizo12} and of non-local Sy 1s from \citet{Woo06, Woo08}. Note also that the Mg II-based black hole masses are systematically lower than the H$\alpha$ based masses. The average \mbh~offset, again excluding SDSS $2306-0100$, based on the Mg II measurements is only 0.64 dex. The small number of objects in our sample preclude us from claiming whether the entire sample of PSQs is on average offset from the local \msig~relation. To determine that, we will need to study more objects from the parent sample of PSQs. Furthermore, it is not unreasonable that for these six objects we will find some that fall above the 1$\sigma$ scatter of the local relation, but it is curious that all of the offsets are on the same side of the relation.

The issue is complicated by the matter of the discrepancy between black hole masses measured with the Mg II and with the H$\alpha$ emission lines. We used the SDSS data to measure the H$\alpha$ emission for all but one object, and the SDSS spectroscopy has low S/N ($<10$ pix$^{-1}$). \citet{Denney09} show that using a Gauss-Hermite polynomial to fit low S/N data may overestimate the FWHM of the H$\beta$ line. In the one object for which we have high S/N LRIS spectroscopy (SDSS $0057+0100$) of the H$\alpha$ region, we measured a lower FWHM of the H$\alpha$ broad line compared to the SDSS spectrum. The difference was $\sim400$ km s$^{-1}$ ($\sim8\%$). The black hole mass for SDSS $0057+0100$ using the larger width is only a factor of 1.2 greater than using the smaller width, which is within our defined errors. Smaller H$\alpha$ widths would bring the H$\alpha$-based mass measurements closer to the Mg II-based measurements. However, this small effect is unlikely to reconcile the two measurements. As an alternative explanation, it is possible there is a systematic difference in BLR gas dynamics between the H$\alpha$ and Mg II-emitting gas.

\subsection{The \mlum~Relation}

\begin{figure*}
\epsscale{1.0}
\plotone{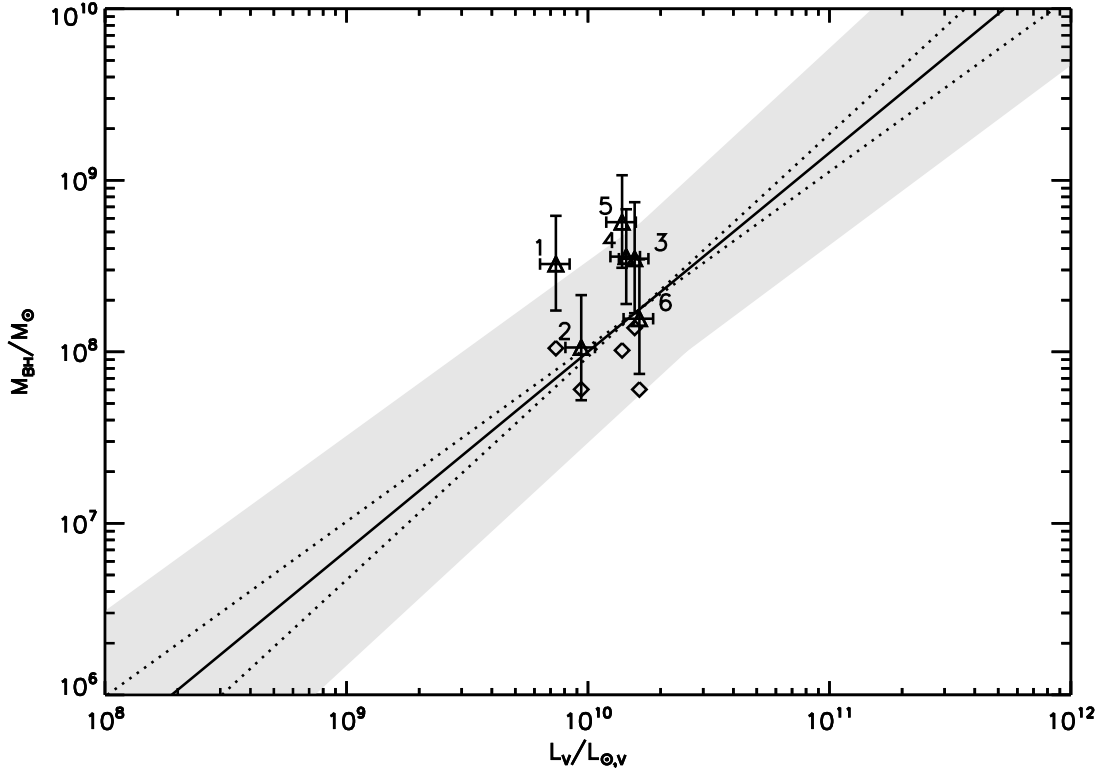}
\caption{The \mlum relation. The solid line is the relation from \citet{McConnell11b}. The dotted lines show the relations defined by the uncertainty in the fitted scale and slope with scatter shown as the grey shaded region ($\pm$0.5 dex). H$\alpha$ based black hole masses of the PSQs are plotted as triangles, and the Mg II based masses of the same objects are plotted as diamonds.\\}
\label{mlum}
\end{figure*}

We plot the \mlum~relation in Fig.~\ref{mlum}. This serves as a test of the bulge properties of the host. If the objects follow this relation, then the black holes of these galaxies may be ``normal'' for their bulges, while the velocity dispersions cause the offset seen in the \msig~relation. In order to compare the bulge luminosities that we have measured here to measurements of systems in the local universe, we need to account for the change in luminosity associated with the evolution of stellar populations.

We measured the bulge luminosities in a procedure that is similar to the measurement of the AGN continuum luminosities. For each object, we scaled our fitted host galaxy template to the measured bulge photometry of \citet{Cales11} in the $HST$ F606W filter. We then measure V band photometry from the scaled host spectrum. Photometry from Cales \etal~has already been k-corrected, but we also apply a passive evolution correction using the online calculator\footnote{www.astro.yale.edu/dokkum/evocalc/} of \citet{vDokkumFranx01}. We applied the ``Default 1: single stellar population'' setting using the calculator, which uses a burst of star formation that passively evolves. This setting approximates the starburst event of our PSQs, which have luminosities dominated by the young population (see Table \ref{sigmas}).

The ``Default 1'' setting of the passive evolution calculator assumes that the stars were formed at redshift $z>>1$. However, the PSQs clearly have had a recent starburst within the past $\le 1$ Gyr. Thus, we adjusted this parameter to more closely match the star formation history of our targets. The difference in redshift from 0.3 (the average of the sample) to 0.4 corresponds to roughly 840 Myr. This is appropriate for all of our targets except SDSS $2118+0056$, which has a redshift closer to 0.4 than 0.3. For this target we adopt the passive evolution correction assuming a formation redshift of 0.5 (a 740 Myr stellar population). As expected, this produced a larger correction than assuming all stars formed at $z>>1$, and applying this correction shifts the objects to smaller V band luminosities. We adopt an error on the bulge luminosities that assumes 0.1 mag error from the fits of Cales \etal~and an additional 10\% error on the spectral host-AGN decomposition.

In Fig.~\ref{mlum}, the solid line shows the best fit relation from \citet{McConnell11b}, who compile measurements of individual galaxies from the literature, including brightest cluster galaxies. The two dotted lines show fits defined by the uncertainty in their scaling and slope, while the grey shaded region shows the intrinsic scatter they measured. Two of the H$\alpha$-based black hole mass measurements (triangles) fall above the scatter in the relation, while the remaining objects are consistent with the relation. However, if we consider the Mg II-based black hole masses (diamonds), all of the objects fall on the relation. We also note that the scatter from McConnell \etal~(0.5 dex) is somewhat greater than that of \citet{Gultekin09}, who found 0.38 dex. If we were to consider the fit from Gultekin \etal, four of the H$\alpha$-based black hole masses would fall above the relation. Thus, to some degree, the interpretation of the offset from the relation depends on the relation's true intrinsic scatter.

The result is quite similar to that from the \msig~relation. That is: PSQs fall within the scatter or above the \mlum~relation if we consider the black hole masses measured from H$\alpha$. Whereas, if we consider the \mbh~measured from Mg II, the PSQs are consistent with the relation. From this we can surmise that if the offset is to be believed, the black holes really are over-massive for their bulges.

\subsection{The Faber-Jackson Relation}

\begin{figure*}
\epsscale{1.0}
\plottwo{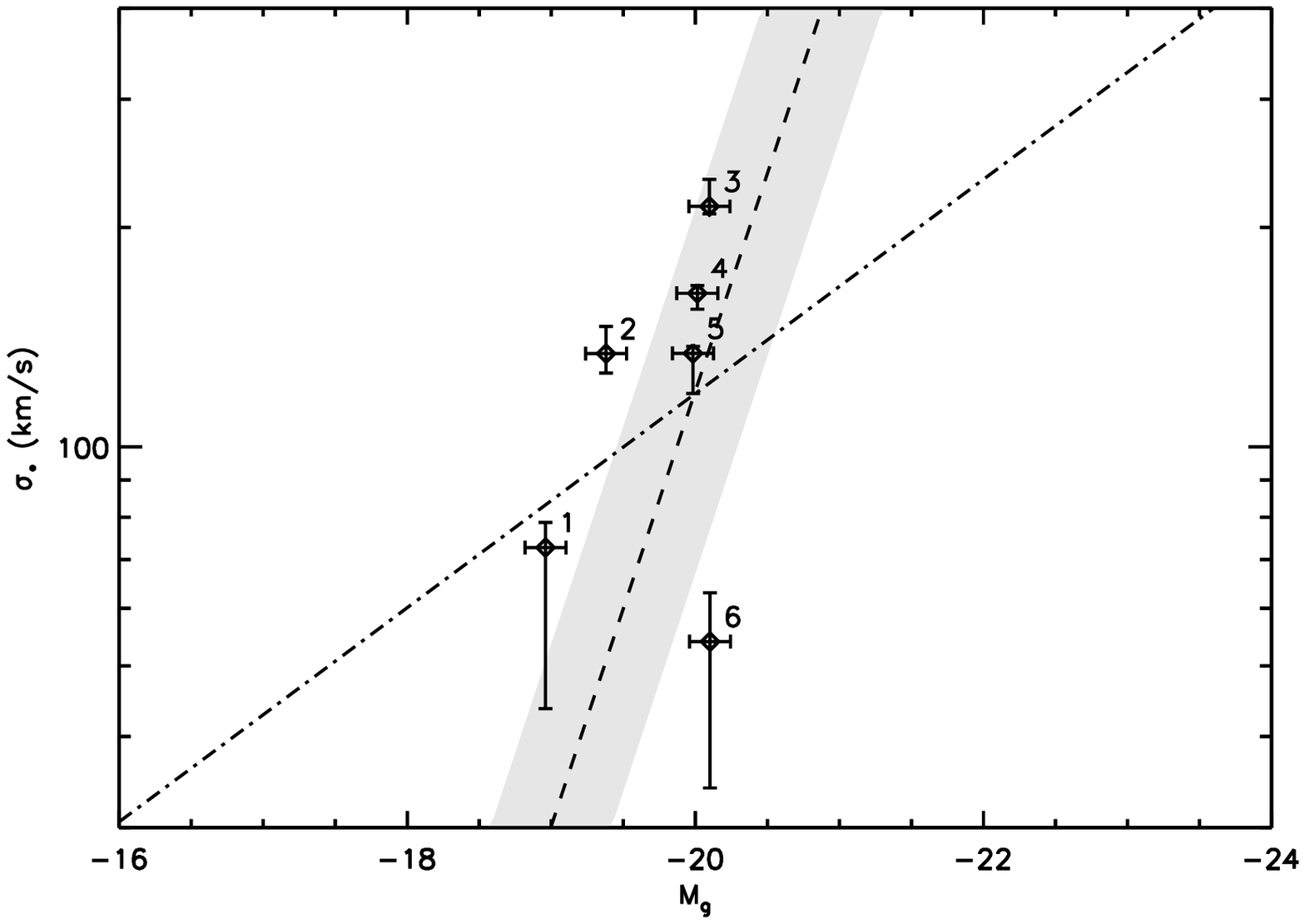}{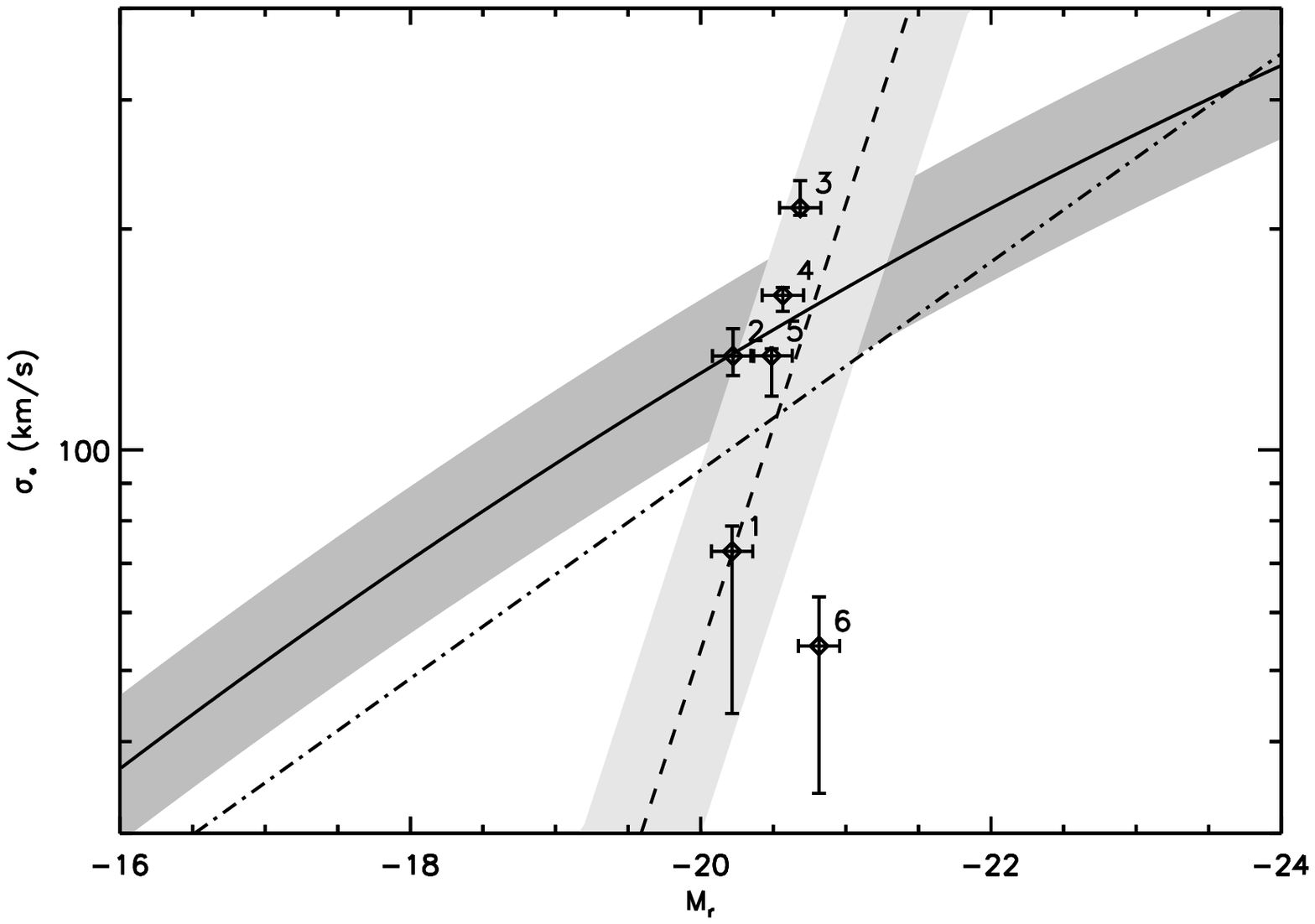}
\caption{The Faber-Jackson relation. {\it Left}: The SDSS $g$ band absolute AB magnitudes. The dashed line is the best fit from \citet{Nigoche10} for the magnitude bin -19.5 $> g >$ -20.5. {\it Right}: The SDSS $r$ band absolute AB magnitudes. The dashed line is the best fit from \citet{Nigoche10} for the magnitude bin -20 $> r >$ -21. The solid line and dark grey region show the quadratic fit and scatter to the Faber-Jackson relation from \citet{Desroches07}. For both panels the the light grey shaded region shows the scatter in the fit from \citet[$\pm$0.25 dex,][]{Nigoche10}, and the dot-dashed line shows the best fit for the overall sample of Nigoche-Netro \etal.\\}
\label{FJR}
\end{figure*}

Figure \ref{FJR} shows the location of our sample on the Faber-Jackson relation \citep{FJ76}. If the \ssig~values that we measured are causing the offset seen from the \msig~relation, we expect the objects to fall below this relation. That is, the objects should have velocity dispersions that are too small for their bulge luminosities. This is the case for one object, but the remaining five are generally consistent with the relation.

The left and right panels show the SDSS $g$ band and $r$ band absolute AB magnitude, respectively. The $g$ and $r$ photometry were measured with the same method that we used to measure the V band luminosity (see previous section) with the additional step of converting luminosities into SDSS AB magnitudes for ease of comparison. This also required transforming the passive evolution corrections from B and V to $g$ and $r$ corrections, for which we used the relations from \citet{Jester05}. \citet{Nigoche10} calculate the Faber-Jackson relation in discrete magnitude bins of width 1 magnitude. Considering the $g$ absolute magnitudes ({\it left panel}), the PSQs appear to follow the relations defined by Nigoche-Netro \etal. The dashed line represents the relation appropriate for objects with magnitudes between $-19.5$ and $-20.5$. Two objects fall in the magnitude bin $-18.5$ to $-19.5$ and are consistent with that relation. The only object that falls significantly offset is SDSS 2306-0100, which is one of the face-on disk galaxies. Considering the $r$ magnitudes ({\it right panel}), all objects fall in the magnitude bin $-20 > M_r > -21$. Five objects are consistent with the relation, while SDSS 2306-0100 is again significantly offset. In both panels we have plotted the overall relation using the entire sample of Nigoche-Netro \etal. The fit for their overall sample has an intrinsic dispersion of 0.6 dex. 

\citet{Desroches07} have also studied the luminosity dependence of the Faber-Jackson relation. They derive a quadratic fit to to the relation in the SDSS $r$ band using data from the NYU Value Added Galaxy Catalog \citep{Blanton05}. While \citet{Nigoche10} use only galaxies that are well fit by a de Vaucouleurs profile and the corresponding magnitudes, \citet{Desroches07} select objects with a range of S\'{e}rsic indices and calculate a ``S\'{e}rsic-like'' magnitude from a Petrosian magnitude. As Desroches \etal~explain, the S\'{e}rsic index is correlated with the galaxy luminosity and therefore may bias the slope of the relation if not accounted for. We have plotted the quadratic fit from Desroches \etal~log \ssig~$= -1.79 + 0.674$(log L)$ - 0.0234$(log L)$^2$; M$ = -2.5$(log L)) with 0.1 dex scatter in the right panel of Fig.~\ref{FJR} (solid line and dark grey shading) along with the fit from \citet{Nigoche10}. Three of our objects fall on the relation derived by Desroches \etal. One object falls above the relation (within 2$\sigma$), while two fall significantly below the relation. The two objects that are significantly offset are also the most offset from the \msig~relation, suggesting that at least part of the offset is due to peculiar dynamics of the stellar population.

\citet{Norton01} studied the velocity dispersions of quiescent E+A galaxies. To separate the post-starburst population from the underlying older population of stars they simultaneously fit the 4100 - 5250 \AA~region with K, G, and A star templates. They find that the post-starburst population does not show a significant correlation between its velocity dispersions and the overall host luminosities. However, the distribution of dispersions for their sample is consistent with the Faber-Jackson relation using data from \citet{Faber89} and \citet{Jorgensen94}. The older stellar population \ssig~does show correlation with host luminosity, but the objects fall on average $\sim$0.6 mag brighter than the elliptical galaxies of similar dispersion. We do not separate the post-starburst population from the older population with unique measures of \ssig~as Norton \etal~do, so the value of \ssig~that we measure has influence from both populations.

\section{Discussion}

We have measured \mbh~and \ssig~in six AGN that are hosted by galaxies showing post-starburst stellar populations (``post-starburst quasars''). These objects fall on or above the locally defined \msig~and \mlum~relations. The location of the PSQs on the \msig~diagram is consistent with other studies of objects at $z>0.1$ \citep{Canalizo12,Woo08,Woo06}. 

We have used the Balmer region, which includes the Balmer series, Ca lines, the G band, and other stellar features to estimate the stellar velocity dispersions. While we have used host galaxy templates that approximate the stellar population of the host galaxies, it is possible that the young and old populations have different velocity dispersions. The kinematics of the young population are influenced by the mechanism that triggered the starburst. In a theoretical work, \citet{Bekki05} describe various models of galaxy interactions (major/minor mergers, strong tidal interactions, etc.) that lead to E+A spectral signatures. They find a range of morphologies, velocity dispersions, and starburst mass fractions resulting from the merger event depending on the specific model used. The kinematic properties of the resulting post-starburst population depend on the initial parameters of the galaxy-galaxy interaction that induces the starburst.

The morphologies presented by \citet{Cales11} for our sample indicate that these objects have only minorly disturbed features, if any, that could indicate a recent merger event. Four objects of our sample are described by Cales \etal~as containing a disk component. However, the overall sample of PSQs display a range of morphologies, and Cales \etal~found that half of the observed sample appear to be post-merger remnants. The variety of morphologies seen in the parent sample is consistent with the simulations conducted by \citet{Bekki05}, indicating a range of possible galaxy-galaxy interactions that trigger the starburst event (\eg~tidal interactions, minor/major mergers). Regardless of the triggering mechanism, the PSQs, at least in this subsample, have \mbh~and \ssig~that are generally consistent with previously studied $z>0.1$ galaxies. 

The origin of the velocity dispersion (pressure supported versus rotational) should influence whether the objects fall on the \msig~relation, which is best defined for pressure supported bulges. Observationally, \citet{Norton01} studied the young and old populations independently in E+A galaxies and found them to be largely pressure supported, with the younger post-starburst population having higher \ssig~than the old population. However, \citet{Bekki05} had trouble reproducing this result with their models. By contrast, \citet{Pracy09} found rotation to play a dominant role in their studied E+As. 

We do not separate the young and old populations in our analysis, so the kinematical differences between the post-starburst population and older population in PSQs remains a question. In principle, we could adopt an additional free parameter in our model to measure the velocity dispersion of the old population separately from the young population. However, this would introduce degeneracies that would be difficult to disentangle, especially with the presence of the active nucleus. We are currently conducting an investigation into the kinematic differences between young and old populations in quiescent E+A galaxies that lack the active nucleus continuum contamination \citep[][in preparation]{Hiner12b}. This analysis will be similar to that of Norton \etal, but we will include multiple wavelength regions. By including longer wavelength regions, such as the \ion{Mg}{1b} and \ion{Ca}{2} triplet regions, we will be able to further distinguish between the young and old stellar populations. 

We have also measured the bulge luminosities of the host galaxies. After plotting the \mlum\ and Faber-Jackson relations, we find a scenario which rules out \ssig~bias as the source of the offset from the \msig~relation for three of the six objects we measured. Two objects (SDSS $0030-1035$ and SDSS $2306-0100$) are significantly offset from the \msig~relation. SDSS $2306-0100$ is additionally offset from the Faber-Jackson relation, indicating that the measured \ssig~may be significantly influenced by the face-on disk of the host galaxy. SDSS $0030-1035$ is consistent with the relation as defined by \citet{Nigoche10}, but it falls significantly below that of \citet{Desroches07}, so it may also be influenced by the face-on disk. While SDSS $0237-0101$ falls above the Desroches \etal~Faber-Jackson relation by less than 2$\sigma$, it is also the object that falls closest to the best fit \msig~relation. A reduction of $\sim50$ km s$^{-1}$ in \ssig~would bring it in line with the Desroches \etal~fit, but drive the object left of the \msig~relation (consistent with the other offsets). Peculiar dynamics can clearly influence the measured velocity dispersion, and with a more statistically significant sample, the general trends will become more apparent.

The offset from the \msig~relation is consistent with that found by \citet{Canalizo12}. Similar to their sample and the samples of \citet{Woo06} and \citet{Woo08}, the black holes measured here all have masses $\ge 10^8$ M$_{\odot}$. This brings up the question of sample selection at high-$z$ ($z>0.1$). \citet{Lauer07} describe a potential bias that selects for high-mass black holes at higher-$z$, which is a product of a steep luminosity function and cosmic scatter in both the \msig~and \mlum~relations. \citet{Canalizo12} also show that highly luminous AGN (log(L$_{5100}$/erg s$^{-1}$) $> 43.6$) are offset from the \msig~relation regardless of their redshifts. All but one of the PSQs in our sample have luminosities in this range. 

We have presented six objects of a larger class of post-starburst quasars \citep[][in preparation]{Brotherton12}. This is clearly not statistically significant and we may have by chance chosen objects that all fall above the \msig~relation. To discuss PSQs as a whole, we require measurements of more objects that define a statistically significant sample. We have obtained data for additional PSQs from the same parent sample that will be exhibited in a future work \citep[][in preparation]{Hiner12c}. In addition, it will be important to disentangle the kinematics of the younger post-starburst population from the underlying older stellar population. For this purpose, we have obtained spectroscopy of quiescent E+A galaxies. We will probe possible differences in \ssig~kinematics between the populations using both the Balmer series and longer wavelength regions (\eg~\ion{Mg}{1b}, CaT).

\acknowledgments
The authors thank the anonymous referee for his/her helpful suggestions. The authors also thank M. Vestergaard for providing the Fe II template, and M. Lacy for providing the quasar composite spectrum used in this work. The authors also thank J.H. Woo for insightful discussions. 

Support for this program was provided by the National Science Foundation, under grant number AST 0507450. Additional support was provided by NASA through a grant from the Space Telescope Science Institute (Program AR-12626), which is operated by the Association of Universities for Research in Astronomy, Incorporated, under NASA contract NAS5-26555. KDH acknowledges support from the visitor program of the Niels Bohr Dark Cosmology Centre, University of Copenhagen and support from the Dean's Dissertation Research Grant provided by the UC Riverside graduate division. 
   
The data presented herein were obtained at the W.M. Keck Observatory, which is operated as a scientific partnership among the California Institute of Technology, the University of California and the National Aeronautics and Space Administration. The Observatory was made possible by the generous financial support of the W.M. Keck Foundation.The authors wish to recognize and acknowledge the very significant cultural role and reverence that the summit of Mauna Kea has always had within the indigenous Hawaiian community.  We are most fortunate to have the opportunity to conduct observations from this mountain.

Funding for the SDSS and SDSS-II has been provided by the Alfred P. Sloan Foundation, the Participating Institutions, the National Science Foundation, the U.S. Department of Energy, the National Aeronautics and Space Administration, the Japanese Monbukagakusho, the Max Planck Society, and the Higher Education Funding Council for England. The SDSS Web Site is http://www.sdss.org/.

The SDSS is managed by the Astrophysical Research Consortium for the Participating Institutions. The Participating Institutions are the American Museum of Natural History, Astrophysical Institute Potsdam, University of Basel, University of Cambridge, Case Western Reserve University, University of Chicago, Drexel University, Fermilab, the Institute for Advanced Study, the Japan Participation Group, Johns Hopkins University, the Joint Institute for Nuclear Astrophysics, the Kavli Institute for Particle Astrophysics and Cosmology, the Korean Scientist Group, the Chinese Academy of Sciences (LAMOST), Los Alamos National Laboratory, the Max-Planck-Institute for Astronomy (MPIA), the Max-Planck-Institute for Astrophysics (MPA), New Mexico State University, Ohio State University, University of Pittsburgh, University of Portsmouth, Princeton University, the United States Naval Observatory, and the University of Washington.

\break

\bibliographystyle{apj}
\bibliography{ref}

\begin{thebibliography}{76}
\expandafter\ifx\csname natexlab\endcsname\relax\def\natexlab#1{#1}\fi

\bibitem[{{Abazajian} {et~al.}(2005){Abazajian}, {Adelman-McCarthy},
  {Ag{\"u}eros}, {Allam}, {Anderson}, {Anderson}, {Annis}, {Bahcall}, {Baldry},
  {Bastian}, {Berlind}, {Bernardi}, {Blanton}, {Bochanski}, {Boroski},
  {Brewington}, {Briggs}, {Brinkmann}, {Brunner}, {Budav{\'a}ri}, {Carey},
  {Castander}, {Connolly}, {Covey}, {Csabai}, {Dalcanton}, {Doi}, {Dong},
  {Eisenstein}, {Evans}, {Fan}, {Finkbeiner}, {Friedman}, {Frieman},
  {Fukugita}, {Gillespie}, {Glazebrook}, {Gray}, {Grebel}, {Gunn}, {Gurbani},
  {Hall}, {Hamabe}, {Harbeck}, {Harris}, {Harris}, {Harvanek}, {Hawley},
  {Hayes}, {Heckman}, {Hendry}, {Hennessy}, {Hindsley}, {Hogan}, {Hogg},
  {Holmgren}, {Holtzman}, {Ichikawa}, {Ichikawa}, {Ivezi{\'c}}, {Jester},
  {Johnston}, {Jorgensen}, {Juri{\'c}}, {Kent}, {Kleinman}, {Knapp}, {Kniazev},
  {Kron}, {Krzesinski}, {Lamb}, {Lampeitl}, {Lee}, {Lin}, {Long}, {Loveday},
  {Lupton}, {Mannery}, {Margon}, {Mart{\'{\i}}nez-Delgado}, {Matsubara},
  {McGehee}, {McKay}, {Meiksin}, {M{\'e}nard}, {Munn}, {Nash}, {Neilsen},
  {Newberg}, {Newman}, {Nichol}, {Nicinski}, {Nieto-Santisteban}, {Nitta},
  {Okamura}, {O'Mullane}, {Owen}, {Padmanabhan}, {Pauls}, {Peoples}, {Pier},
  {Pope}, {Pourbaix}, {Quinn}, {Raddick}, {Richards}, {Richmond}, {Rix},
  {Rockosi}, {Schlegel}, {Schneider}, {Schroeder}, {Scranton}, {Sekiguchi},
  {Sheldon}, {Shimasaku}, {Silvestri}, {Smith}, {Smol{\v c}i{\'c}}, {Snedden},
  {Stebbins}, {Stoughton}, {Strauss}, {SubbaRao}, {Szalay}, {Szapudi},
  {Szkody}, {Szokoly}, {Tegmark}, {Teodoro}, {Thakar}, {Tremonti}, {Tucker},
  {Uomoto}, {Vanden Berk}, {Vandenberg}, {Vogeley}, {Voges}, {Vogt},
  {Walkowicz}, {Wang}, {Weinberg}, {West}, {White}, {Wilhite}, {Xu}, {Yanny},
  {Yasuda}, {Yip}, {Yocum}, {York}, {Zehavi}, {Zibetti}, \&
  {Zucker}}]{Abazajian05}
{Abazajian}, K., {Adelman-McCarthy}, J.~K., {Ag{\"u}eros}, M.~A., {et~al.}
  2005, \aj, 129, 1755

\bibitem[{{Abazajian} {et~al.}(2009){Abazajian}, {Adelman-McCarthy},
  {Ag{\"u}eros}, {Allam}, {Allende Prieto}, {An}, {Anderson}, {Anderson},
  {Annis}, {Bahcall}, \& et~al.}]{Abazajian09}
{Abazajian}, K.~N., {Adelman-McCarthy}, J.~K., {Ag{\"u}eros}, M.~A., {et~al.}
  2009, \apjs, 182, 543

\bibitem[{{Barth} {et~al.}(2002){Barth}, {Ho}, \& {Sargent}}]{Barth02}
{Barth}, A.~J., {Ho}, L.~C., \& {Sargent}, W.~L.~W. 2002, \aj, 124, 2607

\bibitem[{{Barth} {et~al.}(2003){Barth}, {Ho}, \& {Sargent}}]{Barth03}
---. 2003, \apj, 583, 134

\bibitem[{{Beifiori} {et~al.}(2011){Beifiori}, {Courteau}, {Corsini}, \&
  {Zhu}}]{Beifiori11}
{Beifiori}, A., {Courteau}, S., {Corsini}, E.~M., \& {Zhu}, Y. 2011, ArXiv
  e-prints

\bibitem[{{Bekki} {et~al.}(2005){Bekki}, {Couch}, {Shioya}, \&
  {Vazdekis}}]{Bekki05}
{Bekki}, K., {Couch}, W.~J., {Shioya}, Y., \& {Vazdekis}, A. 2005, \mnras, 359,
  949

\bibitem[{{Bennert} {et~al.}(2011){Bennert}, {Auger}, {Treu}, {Woo}, \&
  {Malkan}}]{Bennert11}
{Bennert}, V.~N., {Auger}, M.~W., {Treu}, T., {Woo}, J.-H., \& {Malkan}, M.~A.
  2011, \apj, 726, 59

\bibitem[{{Bentz} {et~al.}(2009{\natexlab{a}}){Bentz}, {Peterson}, {Netzer},
  {Pogge}, \& {Vestergaard}}]{Bentz09a}
{Bentz}, M.~C., {Peterson}, B.~M., {Netzer}, H., {Pogge}, R.~W., \&
  {Vestergaard}, M. 2009{\natexlab{a}}, \apj, 697, 160

\bibitem[{{Bentz} {et~al.}(2009{\natexlab{b}}){Bentz}, {Walsh}, {Barth},
  {Baliber}, {Bennert}, {Canalizo}, {Filippenko}, {Ganeshalingam}, {Gates},
  {Greene}, {Hidas}, {Hiner}, {Lee}, {Li}, {Malkan}, {Minezaki}, {Sakata},
  {Serduke}, {Silverman}, {Steele}, {Stern}, {Street}, {Thornton}, {Treu},
  {Wang}, {Woo}, \& {Yoshii}}]{Bentz09b}
{Bentz}, M.~C., {Walsh}, J.~L., {Barth}, A.~J., {et~al.} 2009{\natexlab{b}},
  \apj, 705, 199

\bibitem[{{Bentz} {et~al.}(2010){Bentz}, {Walsh}, {Barth}, {Yoshii}, {Woo},
  {Wang}, {Treu}, {Thornton}, {Street}, {Steele}, {Silverman}, {Serduke},
  {Sakata}, {Minezaki}, {Malkan}, {Li}, {Lee}, {Hiner}, {Hidas}, {Greene},
  {Gates}, {Ganeshalingam}, {Filippenko}, {Canalizo}, {Bennert}, \&
  {Baliber}}]{Bentz10}
---. 2010, \apj, 716, 993

\bibitem[{{Blanton} {et~al.}(2005){Blanton}, {Schlegel}, {Strauss},
  {Brinkmann}, {Finkbeiner}, {Fukugita}, {Gunn}, {Hogg}, {Ivezi{\'c}}, {Knapp},
  {Lupton}, {Munn}, {Schneider}, {Tegmark}, \& {Zehavi}}]{Blanton05}
{Blanton}, M.~R., {Schlegel}, D.~J., {Strauss}, M.~A., {et~al.} 2005, \aj, 129,
  2562

\bibitem[{{Brotherton} {et~al.}(2010){Brotherton}, {Cales}, {Ganguly}, {Shang},
  {Canalizo}, {Stoll}, {Paul}, \& {Diamond-Stanic}}]{Brotherton10}
{Brotherton}, M., {Cales}, S., {Ganguly}, R., {et~al.} 2010, in IAU Symposium,
  Vol. 267, IAU Symposium, 105--105

\bibitem[{{Brotherton} {et~al.}(2012){Brotherton}, {Paul}, {Ganguly}, {Shang},
  {Fawcett}, {Stoll}, {Diamond-Stanic}, {Canalizo}, \& {Vanden
  Berk}}]{Brotherton12}
{Brotherton}, M., {Paul}, C., {Ganguly}, R., {et~al.} 2012, in preparation

\bibitem[{{Bruzual}(2007)}]{Bruzual07}
{Bruzual}, G. 2007, in Astronomical Society of the Pacific Conference Series,
  Vol. 374, From Stars to Galaxies: Building the Pieces to Build Up the
  Universe, ed. {A.~Vallenari, R.~Tantalo, L.~Portinari, \& A.~Moretti}, 303

\bibitem[{{Cales} {et~al.}(2011){Cales}, {Brotherton}, {Shang}, {Bennert},
  {Canalizo}, {Stoll}, {Ganguly}, {Vanden Berk}, {Paul}, \&
  {Diamond-Stanic}}]{Cales11}
{Cales}, S.~L., {Brotherton}, M.~S., {Shang}, Z., {et~al.} 2011, \apj, 741, 106

\bibitem[{{Cales} {et~al.}(2012){Cales}, {Brotherton}, {Shang}, {Runnoe},
  {DiPompeo}, {Bennert}, {Canalizo}, {Hiner}, {Stoll}, {Ganguly}, \&
  {Diamond-Stanic}}]{Cales12}
---. 2012, in preparation

\bibitem[{{Canalizo} {et~al.}(2012){Canalizo}, {Wold}, {Hiner}, {Lazarova}, \&
  {Lacy}}]{Canalizo12}
{Canalizo}, G., {Wold}, M., {Hiner}, K.~D., {Lazarova}, M., \& {Lacy}, M. 2012,
  submitted

\bibitem[{{Cardelli} {et~al.}(1989){Cardelli}, {Clayton}, \& {Mathis}}]{CCM89}
{Cardelli}, J.~A., {Clayton}, G.~C., \& {Mathis}, J.~S. 1989, \apj, 345, 245

\bibitem[{{Croton}(2006)}]{Croton06}
{Croton}, D.~J. 2006, \mnras, 369, 1808

\bibitem[{{Denney} {et~al.}(2009){Denney}, {Peterson}, {Dietrich},
  {Vestergaard}, \& {Bentz}}]{Denney09}
{Denney}, K.~D., {Peterson}, B.~M., {Dietrich}, M., {Vestergaard}, M., \&
  {Bentz}, M.~C. 2009, \apj, 692, 246

\bibitem[{{Desroches} {et~al.}(2007){Desroches}, {Quataert}, {Ma}, \&
  {West}}]{Desroches07}
{Desroches}, L.-B., {Quataert}, E., {Ma}, C.-P., \& {West}, A.~A. 2007, \mnras,
  377, 402

\bibitem[{{Di Matteo} {et~al.}(2005){Di Matteo}, {Springel}, \&
  {Hernquist}}]{DiMatteo05}
{Di Matteo}, T., {Springel}, V., \& {Hernquist}, L. 2005, \nat, 433, 604

\bibitem[{{Faber} \& {Jackson}(1976)}]{FJ76}
{Faber}, S.~M., \& {Jackson}, R.~E. 1976, \apj, 204, 668

\bibitem[{{Faber} {et~al.}(1989){Faber}, {Wegner}, {Burstein}, {Davies},
  {Dressler}, {Lynden-Bell}, \& {Terlevich}}]{Faber89}
{Faber}, S.~M., {Wegner}, G., {Burstein}, D., {et~al.} 1989, \apjs, 69, 763

\bibitem[{{Ferrarese} \& {Merritt}(2000)}]{FM00}
{Ferrarese}, L., \& {Merritt}, D. 2000, \apjl, 539, L9

\bibitem[{{Fine} {et~al.}(2008){Fine}, {Croom}, {Hopkins}, {Hernquist},
  {Bland-Hawthorn}, {Colless}, {Hall}, {Miller}, {Myers}, {Nichol}, {Pimbblet},
  {Ross}, {Schneider}, {Shanks}, \& {Sharp}}]{Fine08}
{Fine}, S., {Croom}, S.~M., {Hopkins}, P.~F., {et~al.} 2008, \mnras, 390, 1413

\bibitem[{{Gebhardt} {et~al.}(2000){Gebhardt}, {Bender}, {Bower}, {Dressler},
  {Faber}, {Filippenko}, {Green}, {Grillmair}, {Ho}, {Kormendy}, {Lauer},
  {Magorrian}, {Pinkney}, {Richstone}, \& {Tremaine}}]{Gebhardt00}
{Gebhardt}, K., {Bender}, R., {Bower}, G., {et~al.} 2000, \apjl, 539, L13

\bibitem[{{Graham} {et~al.}(2011){Graham}, {Onken}, {Athanassoula}, \&
  {Combes}}]{Graham11}
{Graham}, A.~W., {Onken}, C.~A., {Athanassoula}, E., \& {Combes}, F. 2011,
  \mnras, 412, 2211

\bibitem[{{Greene} \& {Ho}(2005)}]{GH05b}
{Greene}, J.~E., \& {Ho}, L.~C. 2005, \apj, 630, 122

\bibitem[{{Greene} \& {Ho}(2006)}]{GH06b}
---. 2006, \apj, 641, 117

\bibitem[{{Greene} {et~al.}(2010{\natexlab{a}}){Greene}, {Peng}, \&
  {Ludwig}}]{Greene10a}
{Greene}, J.~E., {Peng}, C.~Y., \& {Ludwig}, R.~R. 2010{\natexlab{a}}, \apj,
  709, 937

\bibitem[{{Greene} {et~al.}(2010{\natexlab{b}}){Greene}, {Peng}, {Kim}, {Kuo},
  {Braatz}, {Impellizzeri}, {Condon}, {Lo}, {Henkel}, \& {Reid}}]{Greene10b}
{Greene}, J.~E., {Peng}, C.~Y., {Kim}, M., {et~al.} 2010{\natexlab{b}}, ArXiv
  e-prints

\bibitem[{{G{\"u}ltekin} {et~al.}(2009){G{\"u}ltekin}, {Richstone}, {Gebhardt},
  {Lauer}, {Tremaine}, {Aller}, {Bender}, {Dressler}, {Faber}, {Filippenko},
  {Green}, {Ho}, {Kormendy}, {Magorrian}, {Pinkney}, \& {Siopis}}]{Gultekin09}
{G{\"u}ltekin}, K., {Richstone}, D.~O., {Gebhardt}, K., {et~al.} 2009, \apj,
  698, 198

\bibitem[{{Hiner} \& {Canalizo}(2012)}]{Hiner12b}
{Hiner}, K.~D., \& {Canalizo}, G. 2012, in preparation

\bibitem[{{Hiner} {et~al.}(2012){Hiner}, {Canalizo}, {Brotherton}, {Cales}, \&
  {Wold}}]{Hiner12c}
{Hiner}, K.~D., {Canalizo}, G., {Brotherton}, M., {Cales}, S., \& {Wold}, M.
  2012, in preparation

\bibitem[{{Jester} {et~al.}(2005){Jester}, {Schneider}, {Richards}, {Green},
  {Schmidt}, {Hall}, {Strauss}, {Vanden Berk}, {Stoughton}, {Gunn},
  {Brinkmann}, {Kent}, {Smith}, {Tucker}, \& {Yanny}}]{Jester05}
{Jester}, S., {Schneider}, D.~P., {Richards}, G.~T., {et~al.} 2005, \aj, 130,
  873

\bibitem[{{Jorgensen} \& {Franx}(1994)}]{Jorgensen94}
{Jorgensen}, I., \& {Franx}, M. 1994, \apj, 433, 553

\bibitem[{{J{\o}rgensen} {et~al.}(1999){J{\o}rgensen}, {Franx}, {Hjorth}, \&
  {van Dokkum}}]{Jorgensen99}
{J{\o}rgensen}, I., {Franx}, M., {Hjorth}, J., \& {van Dokkum}, P.~G. 1999,
  \mnras, 308, 833

\bibitem[{{Kaspi} {et~al.}(2000){Kaspi}, {Smith}, {Netzer}, {Maoz}, {Jannuzi},
  \& {Giveon}}]{Kaspi00}
{Kaspi}, S., {Smith}, P.~S., {Netzer}, H., {et~al.} 2000, \apj, 533, 631

\bibitem[{{Kormendy} \& {Richstone}(1995)}]{KR95}
{Kormendy}, J., \& {Richstone}, D. 1995, \araa, 33, 581

\bibitem[{{Kriss}(1994)}]{Kriss94}
{Kriss}, G. 1994, Astronomical Data Analysis Software and Systems, 3, 437

\bibitem[{{Kurk} {et~al.}(2007){Kurk}, {Walter}, {Fan}, {Jiang}, {Riechers},
  {Rix}, {Pentericci}, {Strauss}, {Carilli}, \& {Wagner}}]{Kurk07}
{Kurk}, J.~D., {Walter}, F., {Fan}, X., {et~al.} 2007, \apj, 669, 32

\bibitem[{{Lauer} {et~al.}(2007){Lauer}, {Tremaine}, {Richstone}, \&
  {Faber}}]{Lauer07}
{Lauer}, T.~R., {Tremaine}, S., {Richstone}, D., \& {Faber}, S.~M. 2007, \apj,
  670, 249

\bibitem[{{Magorrian} {et~al.}(1998){Magorrian}, {Tremaine}, {Richstone},
  {Bender}, {Bower}, {Dressler}, {Faber}, {Gebhardt}, {Green}, {Grillmair},
  {Kormendy}, \& {Lauer}}]{Magorrian98}
{Magorrian}, J., {Tremaine}, S., {Richstone}, D., {et~al.} 1998, \aj, 115, 2285

\bibitem[{{Massey} \& {Gronwall}(1990)}]{Massey90}
{Massey}, P., \& {Gronwall}, C. 1990, \apj, 358, 344

\bibitem[{{Massey} {et~al.}(1988){Massey}, {Strobel}, {Barnes}, \&
  {Anderson}}]{Massey88}
{Massey}, P., {Strobel}, K., {Barnes}, J.~V., \& {Anderson}, E. 1988, \apj,
  328, 315

\bibitem[{{McConnell} {et~al.}(2011{\natexlab{a}}){McConnell}, {Ma},
  {Gebhardt}, {Wright}, {Murphy}, {Lauer}, {Graham}, \&
  {Richstone}}]{McConnell11b}
{McConnell}, N.~J., {Ma}, C.-P., {Gebhardt}, K., {et~al.} 2011{\natexlab{a}},
  \nat, 480, 215

\bibitem[{{McConnell} {et~al.}(2011{\natexlab{b}}){McConnell}, {Ma}, {Graham},
  {Gebhardt}, {Lauer}, {Wright}, \& {Richstone}}]{McConnell11a}
{McConnell}, N.~J., {Ma}, C.-P., {Graham}, J.~R., {et~al.} 2011{\natexlab{b}},
  \apj, 728, 100

\bibitem[{{McGill} {et~al.}(2008){McGill}, {Woo}, {Treu}, \&
  {Malkan}}]{McGill08}
{McGill}, K.~L., {Woo}, J.-H., {Treu}, T., \& {Malkan}, M.~A. 2008, \apj, 673,
  703

\bibitem[{{McLure} \& {Dunlop}(2002)}]{McLureDunlop02}
{McLure}, R.~J., \& {Dunlop}, J.~S. 2002, \mnras, 331, 795

\bibitem[{{McLure} \& {Jarvis}(2002)}]{McLure02}
{McLure}, R.~J., \& {Jarvis}, M.~J. 2002, \mnras, 337, 109

\bibitem[{{Nigoche-Netro} {et~al.}(2010){Nigoche-Netro}, {Aguerri}, {Lagos},
  {Ruelas-Mayorga}, {S{\'a}nchez}, \& {Machado}}]{Nigoche10}
{Nigoche-Netro}, A., {Aguerri}, J.~A.~L., {Lagos}, P., {et~al.} 2010, \aap,
  516, A96

\bibitem[{{Norton} {et~al.}(2001){Norton}, {Gebhardt}, {Zabludoff}, \&
  {Zaritsky}}]{Norton01}
{Norton}, S.~A., {Gebhardt}, K., {Zabludoff}, A.~I., \& {Zaritsky}, D. 2001,
  \apj, 557, 150

\bibitem[{{Oke} {et~al.}(1995){Oke}, {Cohen}, {Carr}, {Cromer}, {Dingizian},
  {Harris}, {Labrecque}, {Lucinio}, {Schaal}, {Epps}, \& {Miller}}]{Oke95}
{Oke}, J.~B., {Cohen}, J.~G., {Carr}, M., {et~al.} 1995, \pasp, 107, 375

\bibitem[{{Onken} {et~al.}(2004){Onken}, {Ferrarese}, {Merritt}, {Peterson},
  {Pogge}, {Vestergaard}, \& {Wandel}}]{Onken04}
{Onken}, C.~A., {Ferrarese}, L., {Merritt}, D., {et~al.} 2004, \apj, 615, 645

\bibitem[{{Peng}(2007)}]{Peng07}
{Peng}, C.~Y. 2007, \apj, 671, 1098

\bibitem[{{Pracy} {et~al.}(2009){Pracy}, {Kuntschner}, {Couch}, {Blake},
  {Bekki}, \& {Briggs}}]{Pracy09}
{Pracy}, M.~B., {Kuntschner}, H., {Couch}, W.~J., {et~al.} 2009, \mnras, 396,
  1349

\bibitem[{Press {et~al.}(2007)Press, Teukolsky, Vetterling, \&
  Flannery}]{Press07}
Press, W.~H., Teukolsky, S.~A., Vetterling, W.~T., \& Flannery, B.~P. 2007,
  Numerical Recipes 3rd Edition: The Art of Scientific Computing, 3rd edn. (New
  York, NY, USA: Cambridge University Press)

\bibitem[{{Rafiee} \& {Hall}(2011)}]{Rafiee11}
{Rafiee}, A., \& {Hall}, P.~B. 2011, \apjs, 194, 42

\bibitem[{{Robertson} {et~al.}(2006){Robertson}, {Hernquist}, {Cox}, {Di
  Matteo}, {Hopkins}, {Martini}, \& {Springel}}]{Robertson06}
{Robertson}, B., {Hernquist}, L., {Cox}, T.~J., {et~al.} 2006, \apj, 641, 90

\bibitem[{{Rockosi} {et~al.}(2010){Rockosi}, {Stover}, {Kibrick}, {Lockwood},
  {Peck}, {Cowley}, {Bolte}, {Adkins}, {Alcott}, {Allen}, {Brown}, {Cabak},
  {Deich}, {Hilyard}, {Kassis}, {Lanclos}, {Lewis}, {Pfister}, {Phillips},
  {Robinson}, {Saylor}, {Thompson}, {Ward}, {Wei}, \& {Wright}}]{Rockosi10}
{Rockosi}, C., {Stover}, R., {Kibrick}, R., {et~al.} 2010, in Society of
  Photo-Optical Instrumentation Engineers (SPIE) Conference Series, Vol. 7735,
  Society of Photo-Optical Instrumentation Engineers (SPIE) Conference Series

\bibitem[{{Schlegel} {et~al.}(1998){Schlegel}, {Finkbeiner}, \&
  {Davis}}]{Schlegel98}
{Schlegel}, D.~J., {Finkbeiner}, D.~P., \& {Davis}, M. 1998, \apj, 500, 525

\bibitem[{{Sigut} \& {Pradhan}(2003)}]{Sigut03}
{Sigut}, T.~A.~A., \& {Pradhan}, A.~K. 2003, \apjs, 145, 15

\bibitem[{{van Dokkum}(2001)}]{vDokkum01}
{van Dokkum}, P.~G. 2001, \pasp, 113, 1420

\bibitem[{{van Dokkum} \& {Franx}(2001)}]{vDokkumFranx01}
{van Dokkum}, P.~G., \& {Franx}, M. 2001, \apj, 553, 90

\bibitem[{{Vanden Berk} {et~al.}(2001){Vanden Berk}, {Richards}, {Bauer},
  {Strauss}, {Schneider}, {Heckman}, {York}, {Hall}, {Fan}, {Knapp},
  {Anderson}, {Annis}, {Bahcall}, {Bernardi}, {Briggs}, {Brinkmann}, {Brunner},
  {Burles}, {Carey}, {Castander}, {Connolly}, {Crocker}, {Csabai}, {Doi},
  {Finkbeiner}, {Friedman}, {Frieman}, {Fukugita}, {Gunn}, {Hennessy},
  {Ivezi{\'c}}, {Kent}, {Kunszt}, {Lamb}, {Leger}, {Long}, {Loveday}, {Lupton},
  {Meiksin}, {Merelli}, {Munn}, {Newberg}, {Newcomb}, {Nichol}, {Owen}, {Pier},
  {Pope}, {Rockosi}, {Schlegel}, {Siegmund}, {Smee}, {Snir}, {Stoughton},
  {Stubbs}, {SubbaRao}, {Szalay}, {Szokoly}, {Tremonti}, {Uomoto}, {Waddell},
  {Yanny}, \& {Zheng}}]{VandenBerk01}
{Vanden Berk}, D.~E., {Richards}, G.~T., {Bauer}, A., {et~al.} 2001, \aj, 122,
  549

\bibitem[{{Vestergaard} \& {Osmer}(2009)}]{VO09}
{Vestergaard}, M., \& {Osmer}, P.~S. 2009, \apj, 699, 800

\bibitem[{{Vestergaard} \& {Wilkes}(2001)}]{Vestergaard01}
{Vestergaard}, M., \& {Wilkes}, B.~J. 2001, \apjs, 134, 1

\bibitem[{{Vika} {et~al.}(2011){Vika}, {Driver}, {Cameron}, {Kelvin}, \&
  {Robotham}}]{Vika11}
{Vika}, M., {Driver}, S.~P., {Cameron}, E., {Kelvin}, L., \& {Robotham}, A.
  2011, ArXiv e-prints

\bibitem[{{Walsh} {et~al.}(2009){Walsh}, {Minezaki}, {Bentz}, {Barth},
  {Baliber}, {Li}, {Stern}, {Bennert}, {Brown}, {Canalizo}, {Filippenko},
  {Gates}, {Greene}, {Malkan}, {Sakata}, {Street}, {Treu}, {Woo}, \&
  {Yoshii}}]{Walsh09}
{Walsh}, J.~L., {Minezaki}, T., {Bentz}, M.~C., {et~al.} 2009, \apjs, 185, 156

\bibitem[{{Wold} {et~al.}(2007){Wold}, {Lacy}, \& {Armus}}]{Wold07}
{Wold}, M., {Lacy}, M., \& {Armus}, L. 2007, \aap, 470, 531

\bibitem[{{Woo} {et~al.}(2006){Woo}, {Treu}, {Malkan}, \& {Blandford}}]{Woo06}
{Woo}, J., {Treu}, T., {Malkan}, M.~A., \& {Blandford}, R.~D. 2006, \apj, 645,
  900

\bibitem[{{Woo} {et~al.}(2008){Woo}, {Treu}, {Malkan}, \& {Blandford}}]{Woo08}
---. 2008, \apj, 681, 925

\bibitem[{{Woo} \& {Urry}(2002)}]{WooUrry02}
{Woo}, J.-H., \& {Urry}, C.~M. 2002, \apj, 579, 530

\bibitem[{{Woo} {et~al.}(2010){Woo}, {Treu}, {Barth}, {Wright}, {Walsh},
  {Bentz}, {Martini}, {Bennert}, {Canalizo}, {Filippenko}, {Gates}, {Greene},
  {Li}, {Malkan}, {Stern}, \& {Minezaki}}]{Woo10}
{Woo}, J.-H., {Treu}, T., {Barth}, A.~J., {et~al.} 2010, \apj, 716, 269

\bibitem[{{Xiao} {et~al.}(2011){Xiao}, {Barth}, {Greene}, {Ho}, {Bentz},
  {Ludwig}, \& {Jiang}}]{Xiao11}
{Xiao}, T., {Barth}, A.~J., {Greene}, J.~E., {et~al.} 2011, ArXiv e-prints

\end{thebibliography}

\end{document}